\documentclass[english,aps,prl,notitlepage,twocolumn,superscriptaddress,longbibliography]{revtex4-1}
\pdfoutput=1
\usepackage[T1]{fontenc}
\usepackage[latin9]{inputenc}
\usepackage{xcolor}
\usepackage{pdfcolmk}
\usepackage{mathtools}
\usepackage{amsmath}
\usepackage{graphicx}
\PassOptionsToPackage{normalem}{ulem}
\usepackage{ulem}

\makeatletter

\providecolor{lyxadded}{rgb}{0,0,1}
\providecolor{lyxdeleted}{rgb}{1,0,0}

\DeclareRobustCommand{\lyxsout}[1]{\ifx\\#1\else\sout{#1}\fi}

\usepackage[colorlinks,citecolor=blue,linkcolor=blue]{hyperref}

\makeatother

\usepackage{babel}
\begin{document}
\title{A microscopic theory of Curzon-Ahlborn heat engine}
\author{Y. H. Chen}
\affiliation{School of Physics, Peking University, Beijing, 100871, China}
\author{Jin-Fu Chen}
\address{Beijing Computational Science Research Center, Beijing 100193, China}
\address{Graduate School of China Academy of Engineering Physics, No. 10 Xibeiwang
East Road, Haidian District, Beijing, 100193, China}
\author{Zhaoyu Fei}
\address{Graduate School of China Academy of Engineering Physics, No. 10 Xibeiwang
East Road, Haidian District, Beijing, 100193, China}
\author{H. T. Quan}
\thanks{Corresponding author: htquan@pku.edu.cn}
\affiliation{School of Physics, Peking University, Beijing, 100871, China}
\affiliation{Collaborative Innovation Center of Quantum Matter, Beijing 100871,
China}
\affiliation{Frontiers Science Center for Nano-optoelectronics, Peking University,
Beijing, 100871, China}
\date{\today}
\begin{abstract}
The Curzon-Ahlborn (CA) efficiency, as the efficiency at the maximum
power (EMP) of the endoreversible Carnot engine, has a significant
impact on finite-time thermodynamics. However, the CA engine model
is based on many assumptions. In the past few decades, although a
lot of efforts have been made, a microscopic theory of the CA engine
is still lacking. By adopting the method of the stochastic differential
equation of energy, we formulate a microscopic theory of the CA engine
realized with an underdamped Brownian particle in a class of non-harmonic
potentials. This theory gives microscopic interpretation of all assumptions
made by Curzon and Ahlborn, and thus puts the results about CA engine
on a solid foundation. Also, based on this theory, we obtain analytical
expressions of the power and the efficiency statistics for the Brownian
CA engine. Our research brings new perspectives to experimental studies
of finite-time microscopic heat engines featured with fluctuations.
\end{abstract}
\maketitle
\textit{Introduction.}---For practical heat engines, not only the
efficiency but also the power characterizes the performance. Optimizing
the power and the efficiency of heat-engine cycles is one of the goals
in the study of finite-time thermodynamics \citep{Andresen2011,Berry2020}.
Compared to the Carnot efficiency achieved in infinite time \citep{callen2006thermodynamics},
the efficiency at the maximum power (EMP) attracts a lot of attention
in the studies of finite-time heat engines. The early studies \citep{yvon1955proceedings,chambadal1957recuperation,novikov1958efficiency,curzon1975efficiency,Rubin1979,Moreau2015,Ouerdane2015,Feidt2017}
of the endoreversible Carnot engine concluded that the EMP is the
well-known Curzon-Ahlborn (CA) efficiency
\begin{equation}
\eta_{CA}=1-\sqrt{\frac{T_{C}}{T_{H}}},
\end{equation}
where $T_{C}$ and $T_{H}$ denote the temperatures of the cold and
the hot heat baths. The CA efficiency is in a similar form to the
Carnot efficiency, and is only relevant to the temperatures of the
two heat baths and is independent of any other characteristic of the
heat engine. The CA efficiency aroused a lot of attention and led
to many following-up researches (see for example Refs. \citep{Rubin1979,Rubin1979a,schmiedl2007efficiency,Dechant2017}).
It has become a paradigmatic result in studies dealing with thermodynamic
optimization in the framework of finite-time and stochastic thermodynamics.
In addition, the CA efficiency is relevant to many practical thermal
machines \citep{Esposito_2010}.

The CA efficiency as the EMP has also been derived in some different
setups \citep{Hernandez2014}. For example, in Ref. \citep{van2005thermodynamic}
it is shown that CA efficiency is a result which can be obtained in
the well-founded linear irreversible thermodynamics. In Ref. \citep{Esposito_2010},
the CA efficiency, as the EMP, is derived in the symmetric low-dissipation
regime, where the irreversible entropy production is assumed inversely
proportional to the period of a cycle. Such a $1/\tau$-scaling has
been recently verified in the experiment of finite-time isothermal
compression of dry air \citep{Ma2019,Chen2021}. These studies confirm
the validity of the CA efficiency as the EMP in many heat engine models.
Meanwhile, in some other models, the EMP deviates from the CA efficiency
\citep{van2005thermodynamic,schmiedl2007efficiency,Tu2008,Izumida2008,Esposito_2010,Wang2012,Cavina2017,Ma2018b,Abiuso2020,Abah2012,Cisneros2007,Izumida2009,SanchezSalas2010,Nakamura2020,Fu2020,Lavenda2007,Tu2020,Bauer2016,Chen1994,Chen2019a,Dann2020,Deffner2018,Smith2020,Bonanca2019,Tu2012,Chen2011},
probably due to different circumstances of these models.

The original derivation of the CA efficiency \citep{yvon1955proceedings,chambadal1957recuperation,novikov1958efficiency,curzon1975efficiency}
is based on a lot of assumptions, such as the endoreversible assumption
and the assumption of constant temperature difference. But how reliable
are those assumptions made by Curzon and Ahlborn remains unclarified
due to the lack of a microscopic theory of the CA engine. Also, due
to this lack, the control scheme of the work parameter which is essential
to construct the optimal cycle can not be determined (except for the
harmonic potential \citep{Dechant2017}), neither can the work and
heat statistics as well as the fluctuation theorems of the CA engine.
In the past few decades, a lot of efforts have been made to seek a
microscopic interpretation of the CA engine, but were unsuccessful.

In this Letter, we fill this long-standing gap by realizing the CA
engine with an underdamped Brownian particle \citep{Hondou2000,Blickle2011,Martinez2015,Holubec2020,Gieseler2014,Gong2016,Kwon2013,Dechant2017,GomezMarin2008,Celani2012}
in a time-dependent potential as the working substance. By adopting
the method of stochastic differential equation of energy \citep{Salazar2016,salazar2019stochastic},
we give microscopic interpretation of all assumptions made by Curzon
and Ahlborn, including the endoreversibility, Newton's cooling law
and the constant temperature difference. Thus we lay a solid foundation
for the CA engine model. Furthermore, this microscopic theory allows
us to determine the control scheme of the work parameter of the Brownian
CA engine and study the fluctuations of the power and the efficiency
of the Brownian CA engine. Our study demonstrates that when downsizing
the working substance to a single Brownian particle, results about
the average power and efficiency of the CA engine remain valid, but
the fluctuations become prominent.

\textit{The model.}---The working substance of the engine is modeled
as a Brownian particle\textbf{ }\citep{Jun2014,Gieseler2014,Martinez2015,Martinez2016,Martinez2017,Hoang2018,Albay2020,Tu2014,Li2017,Li2019,Paneru2018,Speck2011,Schmiedl2007,Rana2014}
constrained in a controllable potential $\mathcal{U}(x,t)=k(t)x^{2n}/(2n)$
with the control parameter $k(t)$ and a positive integer $n$. In
the isothermal expansion (compression) process, the control parameter
$k(t)$ is varied when the engine is in contact with the heat bath
at temperature $T_{\mathrm{b}}$. The motion of the particle with
mass $m$ is governed by the complete Langevin equation

\begin{equation}
\ddot{x}+\gamma\dot{x}+k(t)x^{2n-1}=\frac{1}{m}\xi(t),
\end{equation}
where the random force $\xi(t)$ represents a Gaussian white noise
satisfying $\langle\xi(t)\rangle=0$ and $\langle\xi(t)\xi(t')\rangle=2m\gamma T_{\mathrm{b}}\delta(t-t')$
with the friction coefficient $\gamma$. Throughout the text the Boltzmann
constant is set to be $k_{B}=1$.

We consider the highly underdamped regime $\tau_{p}\ll\gamma^{-1}$
and slow external driving $\tau_{p}\ll k/\dot{k}$, where $\tau_{p}$
is the period of the unperturbed motion of the particle, e.g., $\tau_{p}=2\pi\sqrt{m/k}$
for a harmonic oscillator ($n=1$). Under these two conditions, the
variation of the stochastic energy of the particle within a period
is relatively small, which allows us to study the dynamics of the
stochastic energy $E=m\dot{x}^{2}/2+\mathcal{U}(x,t)$. Based on Ito's
lemma and Virial theorem, the equation of motion is expressed as the
stochastic differential equation of the energy $E$ \citep{Salazar2016,salazar2019stochastic}

\begin{equation}
dE=\frac{\dot{\lambda}}{\lambda}Edt-\Gamma\left(E-\frac{f_{n}T_{\mathrm{b}}}{2}\right)dt+\sqrt{2\Gamma T_{\mathrm{b}}E}dB_{t}.\label{eq:dE_eq}
\end{equation}
where the work parameter is rewritten into $\lambda(t)=k(t)^{1/(n+1)}$
with the increment $dB_{t}$ of the Wiener process, the effective
friction coefficient $\Gamma=2n\gamma/(n+1)$, and the effective degrees
of freedom $f_{n}=1+1/n$. The increment of trajectory work for this
system is
\begin{equation}
dW=\frac{\dot{\lambda}}{\lambda}Edt.\label{stochastic_WORK}
\end{equation}
The trajectory heat is obtained from the first law of thermodynamics
as

\begin{equation}
dQ=-\Gamma\left(E-\frac{f_{n}T_{\mathrm{b}}}{2}\right)dt+\sqrt{2\Gamma T_{\mathrm{b}}E}dB_{t}.\label{eq:stochastic_heat}
\end{equation}

The Fokker-Planck equation associated with Eq. (\ref{eq:dE_eq}) can
be solved explicitly \citep{salazar2019stochastic}. During the dynamical
evolution process, the system remains in a Maxwell-Boltzmann distribution
in the energy space and thus can be described by an effective temperature
$\theta(t)$ as

\begin{equation}
P(E,t)=\frac{e^{-E/\theta(t)}}{\Gamma(f_{n}/2)}\frac{E^{f_{n}/2-1}}{\theta(t)^{f_{n}/2}},\label{eq:MB}
\end{equation}
where $\Gamma(x)=\int_{0}^{\infty}e^{-y}y^{x-1}dy$ is gamma function.
Notice that Eq. (\ref{eq:MB}) leads to the endoreversibility, which
is usually assumed in previous studies relevant to the CA engine,
but is derived as a consequence of the equation of motion in our setup.
The ensemble average of the energy is $\langle E(t)\rangle=f_{n}\theta(t)/2$
with the effective temperature $\theta(t)$ governed by 
\begin{equation}
\dot{\theta}(t)=\frac{\dot{\lambda}}{\text{\ensuremath{\lambda}}}\theta(t)-\Gamma[\theta(t)-T_{\mathrm{b}}].\label{eq:dotthetatemperature}
\end{equation}

We emphasize that the l.h.s. of Eq. (\ref{eq:dotthetatemperature})
corresponds to the time derivative of the average energy up to a factor
$f_{n}/2$. The two terms on the r.h.s. correspond to the average
work flux and heat flux, respectively. The average heat flux satisfies
Newton's cooling law, which is also derived as a consequence of the
equation of motion. The effective friction coefficient $\Gamma$,
as a cooling rate, is independent of the work parameter $\lambda$.
We would like to point out that a similar equation of motion for the
effective temperature $\theta(t)$ has been obtained previously for
the ideal gas as the working substance \citep{Rubin1979a}. However,
their derivation relies on several assumptions, for example, the equation
of state of ideal gas, the phenomenological Newton's cooling law and
the endoreversible assumption. On the contrary, the results presented
here are all derived from the microscopic dynamics, and are capable
of describing microscopic systems featured with fluctuations.

\global\long\def\theenumi{\roman{enumi}}%
 
\global\long\def\labelenumi{(\theenumi)}%

\textit{Realization of Curzon-Ahlborn engine based on a Brownian particle.}---With
the model introduced above, we study the EMP of such a microscopic
Brownian engine and formulate a microscopic theory of the CA engine.
To construct a finite-time Carnot cycle, two heat baths at different
temperatures $T_{i},\:i=H,C$ are required in the hot and the cold
isothermal processes. The (effective) friction coefficients $\gamma_{i}$
($\Gamma_{i}$) may be different in the two processes. Based on Curzon
and Ahlborn's derivation \citep{curzon1975efficiency}, we summarize
the preconditions of the CA engine as follows
\begin{enumerate}
\item \label{item1} \textbf{Endoreversibility }\citep{Hoffmann1997}\textbf{.}
The state of the working substance of the engine can be described
by an effective temperature.
\item \label{item2} \textbf{Newton's cooling law (or linear heat transfer
law).} The heat flux between the working substance and the heat bath
is proportional to the temperature difference.
\item \label{item3} \textbf{Constant temperature difference.} During the
isothermal expansion (compression) process, the effective temperature
of the working substance remains at a constant value $\theta_{H}\:(\theta_{C})$
different from that of the heat bath $T_{H}\,(T_{C})$.
\item \label{item4} \textbf{Internal reversible Carnot cycle.} All irreversibilities
are associated with the heat exchange between the working substance
and the heat baths while the adiabatic processes remain reversible
\citep{Hernandez2014}. The heat engine operates like a reversible
Carnot engine between two virtual heat baths at temperatures $\theta_{C}$
and $\theta_{H}$, respectively.
\item \label{item5} \textbf{Constant heat capacity }\citep{Johal2021}\textbf{
and cooling rate} \citep{Apertet2017}\textbf{. }Both the heat capacity
of the working substance and the cooling rate are independent of the
temperature and the work parameters.
\end{enumerate}
According to Eqs. (\ref{eq:MB}) and (\ref{eq:dotthetatemperature}),
our setup fulfills requirements (\ref{item1}) and (\ref{item2}).
In supplementary material, we prove that the optimal cycle corresponding
to the maximum power is exactly the CA cycle satisfying preconditions
(\ref{item3}) and (\ref{item4}). Precondition (\ref{item5}) is
guaranteed by the generalized equipartition theorem and the highly
underdamped condition.

\begin{figure}[htbp]
\begin{centering}
\includegraphics[width=1\columnwidth]{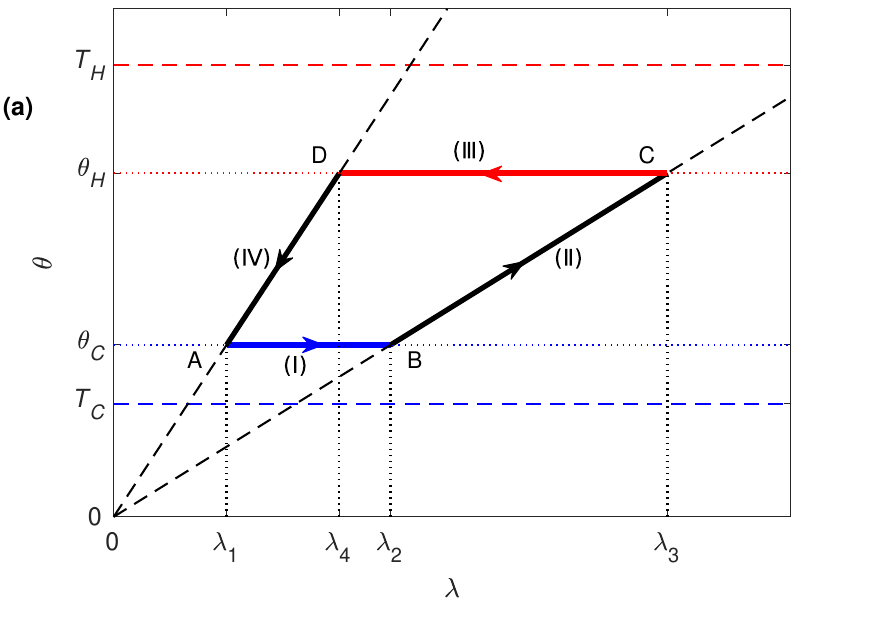}
\par\end{centering}
\begin{centering}
\includegraphics[width=1\columnwidth]{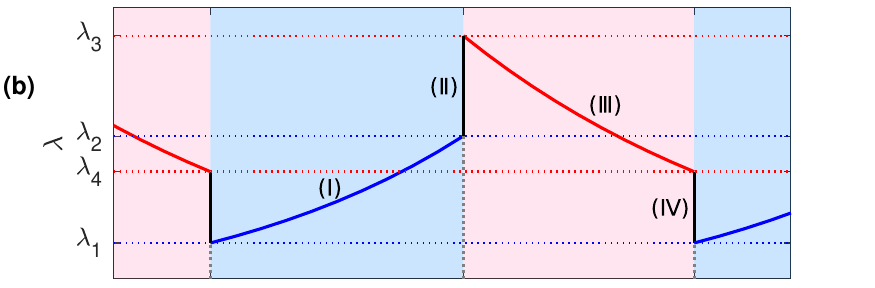}
\par\end{centering}
\begin{centering}
\includegraphics[width=1\columnwidth]{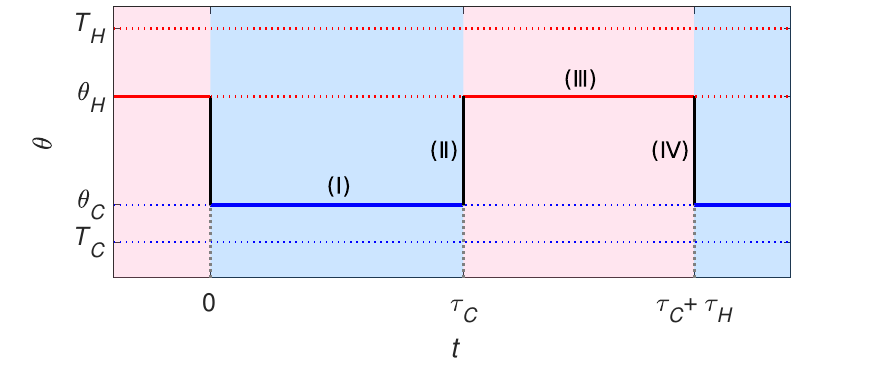}
\par\end{centering}
\centering{}\caption{CA cycle based on a Brownian particle. (a) The cycle diagram in the
space of the work parameter $\lambda$ and the effective temperature
$\theta$. In the isothermal expansion (compression) process, the
working substance remains at a constant effective temperature $\theta_{H}(\theta_{C})$,
which is different from the temperature $T_{H}(T_{C})$ of the hot
(cold) heat bath. In the two adiabatic processes, the effective temperature
$\theta$ of the working substance is proportional to the work parameter
$\lambda$. (b) The control scheme of the work parameter $\lambda(t)$
and the evolution of the effective temperature $\theta(t)$ in a finite-time
cycle. The work parameter $\lambda$ is varied exponentially with
time in an isothermal process, and is quenched abruptly in an adiabatic
process.\label{fig1}}
\end{figure}

Based on the equation of motion of the effective temperature (Eq.
(\ref{eq:dotthetatemperature})) and the definition of trajectory
work and heat, we can optimize the average power of the finite-time
Brownian engine. In Fig. \ref{fig1}(a), we plot the cycle diagram
of the Brownian CA engine. In order to construct a closed CA cycle,
the values of work parameter at the end of each process in Fig. \ref{fig1}(a)
satisfy
\begin{equation}
\frac{\lambda_{2}}{\lambda_{1}}=\frac{\lambda_{3}}{\lambda_{4}}=r.\label{eq:closedcycle}
\end{equation}
The four processes of the CA cycle are illustrated as follows (see
supplementary material),
\global\long\def\theenumi{\Roman{enumi}}%
 
\begin{enumerate}
\item \label{itema}Isothermal compression. The working substance is in
contact with the cold heat bath. Initiated from $\lambda_{1}$ at
$t=0$, the work parameter is varied exponentially with time $\lambda(t)=\lambda_{1}(\lambda_{2}/\lambda_{1})^{t/\tau_{C}}$,
where $\tau_{C}$ is the duration of the process. From the protocol,
it can be found that the effective temperature of the working substance
remains at a constant during the process
\begin{equation}
\theta_{C}=\frac{\tau_{C}\Gamma_{C}}{\tau_{C}\Gamma_{C}-\ln r}T_{C}.\label{eq:workingfluidtemperature}
\end{equation}
The heat released to the cold heat bath during the isothermal compression
process is

\begin{equation}
-\left\langle Q_{C}\right\rangle =\frac{f_{n}}{2}\Gamma_{C}(\theta_{C}-T_{C})\tau_{C}.\label{eq:coldisothermal_heat}
\end{equation}

\item \label{itemb}Adiabatic compression. The work parameter $\lambda(t)$
is quenched instantaneously from $\lambda_{2}$ to $\lambda_{3}$
with the effective temperature $\theta(t)$ changing from $\theta_{C}$
to $\theta_{H}$ accordingly. When the timescale of varying the work
parameter is much shorter than that of the heat dissipation, Eq. (\ref{eq:dotthetatemperature})
becomes $\dot{\theta}(t)=\dot{\lambda}\theta(t)/\ensuremath{\lambda}$,
which leads to $\theta_{H}/\theta_{C}=\lambda_{3}/\lambda_{2}$.
\item \label{itemc}Isothermal expansion. The working substance is in contact
with the hot heat bath. The work parameter is varied exponentially
with time $\lambda(t)=\lambda_{3}(\lambda_{4}/\lambda_{3})^{(t-\tau_{C})/\tau_{H}}$,
and the working substance remains at a constant effective temperature
\begin{equation}
\theta_{H}=\frac{\tau_{H}\Gamma_{H}}{\tau_{H}\Gamma_{H}+\ln r}T_{H},\label{eq:workingfiulrtmeperature_hot}
\end{equation}
where $\tau_{H}$ is the duration of the process. The heat absorbed
from the hot heat bath during the isothermal expansion process is
\begin{equation}
\left\langle Q_{H}\right\rangle =\frac{f_{n}}{2}\Gamma_{H}(T_{H}-\theta_{H})\tau_{H}.\label{eq:hotisothermal_heat}
\end{equation}
\item \label{itemd}Adiabatic expansion. Finally, the work parameter is
quenched from $\lambda_{4}$ to the initial value $\lambda_{1}$ instantaneously
with the effective temperature changing from $\theta_{H}$ to $\theta_{C}$
accordingly, satisfying $\theta_{H}/\theta_{C}=\lambda_{4}/\lambda_{1}$.
\end{enumerate}
The control scheme of the work parameter $\lambda$ and the evolution
of the effective temperature $\theta$ in a finite-time cycle are
illustrated in Fig. \ref{fig1}(b).

Combing Eqs. (\ref{eq:workingfluidtemperature})-(\ref{eq:hotisothermal_heat}),
it is straightforward to verify the precondition (\ref{item4}) that
the entropy change of the working substance after a cycle is zero
$\Delta S=\left\langle Q_{H}\right\rangle /\theta_{H}+\left\langle Q_{C}\right\rangle /\theta_{C}=0.$
Therefore, the microscopic dynamics of the model, together with the
explicit control scheme $\lambda(t)$, constitutes a microscopic theory
of the CA engine.

The net work of a full cycle is $-\left\langle W\right\rangle =\left\langle Q_{H}\right\rangle +\left\langle Q_{C}\right\rangle $,
and the average power and the average efficiency follow as $\overline{P}\coloneqq-\left\langle W\right\rangle /(\tau_{H}+\tau_{C})$
and $\eta\coloneqq-\left\langle W\right\rangle /\left\langle Q_{H}\right\rangle $,
which are explicitly

\begin{equation}
\overline{P}=\frac{f_{n}\ln r}{2(\tau_{H}+\tau_{C})}\left(\frac{\tau_{H}\Gamma_{H}T_{H}}{\tau_{H}\Gamma_{H}+\ln r}-\frac{\tau_{C}\Gamma_{C}T_{C}}{\tau_{C}\Gamma_{C}-\ln r}\right),\label{eq:power}
\end{equation}
and

\begin{equation}
\eta=1-\frac{1+(\tau_{H}\Gamma_{H})^{-1}\ln r}{1-(\tau_{C}\Gamma_{C})^{-1}\ln r}\frac{T_{C}}{T_{H}}.\label{eq:efficiency}
\end{equation}

In order to achieve the maximum power, we first fix $r$ and optimize
the power over $\tau_{H}$ and $\tau_{C}$. The maximum power is obtained
as (see supplementary material)
\begin{equation}
\overline{P}_{\mathrm{max}}=\frac{f_{n}\Gamma_{C}\Gamma_{H}\left(\sqrt{T_{H}}-\sqrt{T_{C}}\right)^{2}}{2\left(\sqrt{\Gamma_{H}}+\sqrt{\Gamma_{C}}\right)^{2}},\label{eq:max_power}
\end{equation}
 with the corresponding optimal duration of the two isothermal processes
\begin{align}
 & \tau_{H}^{\mathrm{max}}=\frac{\ln r\left(\sqrt{\Gamma_{H}T_{H}}+\sqrt{\Gamma_{C}T_{C}}\right)}{\Gamma_{H}\sqrt{\Gamma_{C}}\left(\sqrt{T_{H}}-\sqrt{T_{C}}\right)},\nonumber \\
 & \tau_{C}^{\mathrm{max}}=\frac{\ln r\left(\sqrt{\Gamma_{H}T_{H}}+\sqrt{\Gamma_{C}T_{C}}\right)}{\Gamma_{C}\sqrt{\Gamma_{H}}\left(\sqrt{T_{H}}-\sqrt{T_{C}}\right)}.\label{eq:ca_time}
\end{align}
Please note that $\overline{P}_{\mathrm{max}}$ is independent of
$r$. Hence $\overline{P}_{\mathrm{max}}$ is also the global maximum
power. It is straightforward to see that the EMP of the Brownian engine
in the highly underdamped regime is the CA efficiency 
\begin{equation}
\eta_{\mathrm{EMP}}=1-\sqrt{\frac{T_{C}}{T_{H}}}=\eta_{CA},
\end{equation}
as we expect. Based on Eqs. (\ref{eq:efficiency}) and (\ref{eq:max_power}),
we derive the trade-off relation between power and efficiency in supplementary
material. Compared to previous studies \citep{shiraishi2016universal,Holubec2016,GonzalezAyala2017,pietzonka2018universal,Ma2018,Abiuso2020},
our trade-off relation is tight and is shown to be reachable with
the explicit control scheme of the work parameter $\lambda(t)$. It
is worth mentioning that Ref. \citep{Chen1989} obtains the same tight
trade-off relation, and Ref. \citep{Dechant2017} obtains the tight
trade-off relation as well as the control scheme for the harmonic
potential. As a generalization, our results are valid for a Brownian
particle in a class of non-harmonic potentials.

\textit{Generating function of work and heat in a finite-time isothermal
process.}---For a microscopic Brownian engine, average values are
insufficient to characterize the performance. Fluctuations are non-negligible
\citep{Holubec2021}. To evaluate the performance of a finite-time
heat engine, we need to quantify the extracted work and heat absorbed
from the hot bath in one heat-engine cycle. For the dynamics described
by the above model, we can derive the analytical results of the joint
generating function of work and heat $I(u,s)\coloneqq\left\langle e^{uQ+sW}\right\rangle $
by generalizing the techniques used in Ref. \citep{salazar2020work}.
The result is

\begin{equation}
I(u,s)=\left[\frac{1+u\tilde{\psi}(\tau)}{1+u\tilde{\psi}_{0}}\right]^{\frac{f_{n}}{2}}e^{\frac{f_{n}(s-u)}{2}\int_{0}^{\tau}\frac{\dot{\lambda}(t)}{\lambda(t)}\psi(t)dt},\label{eq:I(u,s)}
\end{equation}
where $\tau$ is the time duration of the process, and the temperature-like
variable $\psi(t)$ satisfies

\begin{equation}
\frac{d\psi}{dt}=\frac{\dot{\lambda}}{\text{\ensuremath{\lambda}}}\psi-\Gamma(\psi-T_{\mathrm{b}})+(s-u)\frac{\dot{\lambda}}{\lambda}\psi^{2}.\label{eq:continuumlimitdifferential_equation}
\end{equation}
with the initial condition $\psi(0)=\tilde{\psi}_{0}/(1+u\tilde{\psi}_{0})$.
The initial value $\tilde{\psi}_{0}$ is either set as the initial
temperature $\theta_{0}$ or obtained from the previous process. A
shifted temperature-like variable is defined as $\tilde{\psi}(t)\coloneqq\psi(t)/[1-u\psi(t)]$,
whose value $\tilde{\psi}(\tau)$ at the end of this process is used
as the initial value $\tilde{\psi}_{0}$ of the subsequent process.
Detailed derivations to Eqs. (\ref{eq:I(u,s)}) and (\ref{eq:continuumlimitdifferential_equation})
and the analytical expression of the joint generating function $I(u,s)$
are left in supplementary material.

\textit{Statistics of power and efficiency of the Brownian CA engine.}---Based
on the microscopic theory, especially the joint generating function
of work and heat $I(u,s)$ and the control scheme $\lambda(t)$ of
the full cycle, we can further study the fluctuations of the power
and the efficiency \citep{Holubec2014,Denzler2021,Denzler2020,Denzler2020a,Holubec2021,salazar2020work}
together with the fluctuation theorems \citep{sinitsyn2011fluctuation,seifert2012stochastic}
of the finite-time Brownian Carnot engine. Specifically, we calculate
the distribution $p(P)$ of the fluctuating power $P\coloneqq-W/(\tau_{C}+\tau_{H})$
and the distribution $p(\zeta)$ of the fluctuating efficiency $\zeta\coloneqq-(W+\eta Q_{H})/\left\langle Q_{H}\right\rangle $
from the generating function $I_{\mathrm{cycle}}(u_{H},u_{C},s)=\left\langle e^{u_{H}Q_{H}+u_{C}Q_{C}+sW}\right\rangle $
(see supplementary material) of a whole cycle. Please note that, instead
of $-W/Q_{H}$, we define $\zeta$ as the fluctuating efficiency which
characterizes the deviation from the average efficiency $\eta$ defined
above. It is straightforward to see that $\left\langle \zeta\right\rangle =0$.

\begin{figure}[htbp]
\begin{centering}
\includegraphics[width=1\columnwidth]{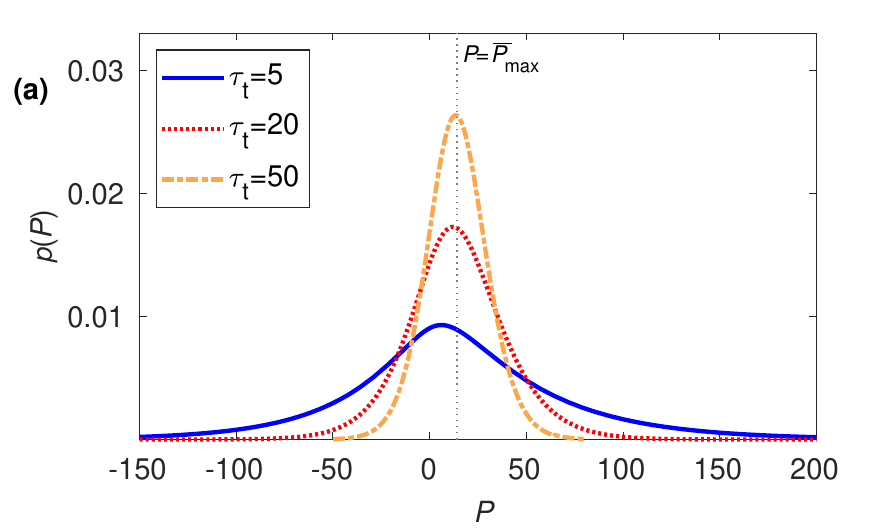}
\par\end{centering}
\begin{centering}
\includegraphics[width=1\columnwidth]{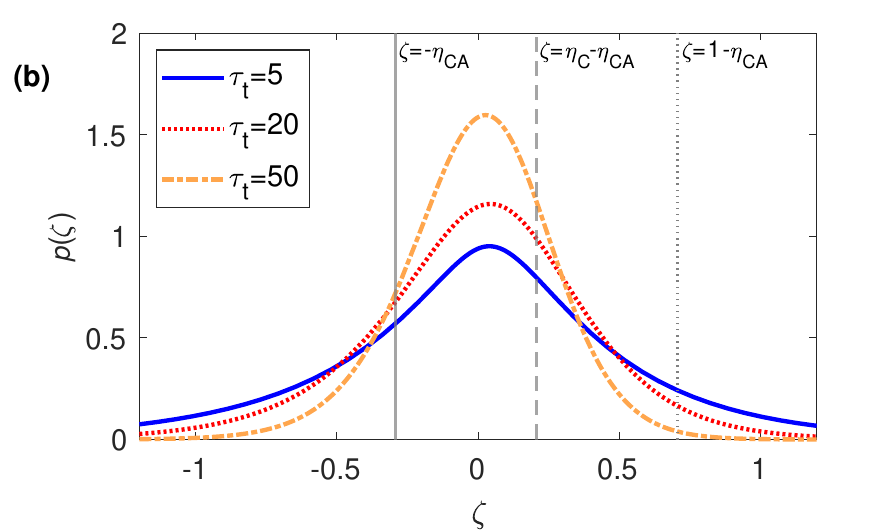}
\par\end{centering}
\centering{}\caption{Distribution of the power (a) and the efficiency (b) of a Brownian
CA engine for three different periods $\tau_{\mathrm{t}}=5,\,20,\,50$.
(a) The vertical dotted line indicates the average power. (b) The
vertical solid, dashed and dotted lines correspond to efficiency $\eta=0$,
$\eta=\eta_{C}$ (Carnot efficiency) and $\eta=1$ respectively. Here
we have chosen $T_{C}=300$, $T_{H}=600$, $\Gamma_{C}=1$, $\Gamma_{H}=1.2$.
\label{fig2}}
\end{figure}

As a special case, we plot the distributions of the power and the
efficiency of the Brownian CA engine in Fig. \ref{fig2}, where $\eta=\eta_{CA}$,
$\tau_{C}=\tau_{C}^{\mathrm{max}}$, $\tau_{H}=\tau_{H}^{\mathrm{max}}$.
Due to the fluctuation, the power can be negative or much larger than
the average power. Similarly, the efficiency can be negative or larger
than Carnot efficiency (even larger than unity). From the analytical
results of the joint generating function $I_{\mathrm{cycle}}(u_{H},u_{C},s)$,
we can show the tendency of the distributions when we increase the
duration of the cycle $\tau_{\mathrm{t}}\coloneqq\tau_{C}^{\mathrm{max}}+\tau_{H}^{\mathrm{max}}$
by increasing $\ln r$
\begin{align}
\mathrm{Var}(P) & \approx\frac{4\overline{P}_{\mathrm{max}}^{2}}{f_{n}\eta_{CA}^{2}}\frac{\left[(1-\eta_{CA})^{2}+1/\delta\right]\left(1+\delta\right)}{\sqrt{\Gamma_{C}\Gamma_{H}}}\frac{1}{\tau_{\mathrm{t}}},\\
\mathrm{Var}(\zeta) & \approx\frac{4(1-\eta_{CA})^{2}}{f_{n}\sqrt{\Gamma_{C}\Gamma_{H}}}\frac{(1+\delta)^{2}}{\delta}\frac{1}{\tau_{\mathrm{t}}},
\end{align}
where $\delta=\sqrt{\Gamma_{H}T_{H}/(\Gamma_{C}T_{C})}$. For both
the power and the efficiency, their variances decrease inversely with
$\tau_{\mathrm{t}}$.

\textit{Summary and discussion.}---In this Letter, we realize the
Curzon-Ahlborn heat engine with a Brownian particle in the highly
underdamped regime. By adopting the method of stochastic differential
equation of energy, we formulate a microscopic theory of the CA engine
based on this model. This theory gives microscopic interpretation
of all assumptions of the CA engine model including the endoreversibility,
Newton's cooling law and the constant temperature difference. Hence,
we lay a solid foundation for the CA engine.

From this microscopic theory, the explicit control scheme $\lambda(t)$
of the CA engine can be uniquely determined, which leads to the maximum
power of the Brownian engine. The control scheme associated with the
maximum power for any given efficiency can be obtained based on the
microscopic theory. In addition, we calculate the generating function
of work and heat, and obtain the analytical results of statistics
of the power and the efficiency together with the fluctuation theorems
of the Brownian CA engine. These quantitative results about the CA
engine bring important insights to the studies of finite-time thermodynamics
beyond the low-dissipation regime \citep{Esposito_2010,Holubec2016,GonzalezAyala2017,Ma2018,GonzalezAyala2020,Abiuso2020,Ma2020}.
For example, results about the average power and efficiency of the
CA engine remain valid when downsizing the working substance to a
single Brownian particle, but fluctuations become prominent. Our study
will shed new light on the experimental explorations about finite-time
Brownian engine, and may inspire future studies about the design of
nanomachines with higher power and efficiency.

\textit{Acknowledgement.}---H. T. Quan thanks Zhan-Chun Tu for valuable
comments and acknowledges support from the National Science Foundation
of China under Grants No. 11775001, No. 11534002, and No. 11825001.

\bibliographystyle{apsrev4-1}
\bibliography{ref,add_reference}

\end{document}


\title{Supplementary Materials: A microscopic theory of Curzon-Ahlborn heat
engine}
\author{Y. H. Chen}
\affiliation{School of Physics, Peking University, Beijing, 100871, China}
\author{Jin-Fu Chen}
\address{Beijing Computational Science Research Center, Beijing 100193, China}
\address{Graduate School of China Academy of Engineering Physics, No. 10 Xibeiwang
East Road, Haidian District, Beijing, 100193, China}
\author{Zhaoyu Fei}
\address{Graduate School of China Academy of Engineering Physics, No. 10 Xibeiwang
East Road, Haidian District, Beijing, 100193, China}
\author{H. T. Quan}
\thanks{Corresponding author: htquan@pku.edu.cn}
\affiliation{School of Physics, Peking University, Beijing, 100871, China}
\affiliation{Collaborative Innovation Center of Quantum Matter, Beijing 100871,
China}
\affiliation{Frontiers Science Center for Nano-optoelectronics, Peking University,
Beijing, 100871, China}
\date{\today}

\maketitle
The supplementary materials are organized as follows. In Sec. \ref{sec:Construction-and-optimization},
we construct and optimize the finite-time Carnot engine with an underdamped
Brownian particle as the working substance, and obtain the power and
the efficiency of the finite-time Brownian engine. In Sec. \ref{sec:tradeoffrelation},
we derive the trade-off relation between the power and the efficiency
for the finite-time Brownian Carnot engine. In Sec. \ref{sec:Calculation-of-the-1},
we calculate the joint generating function of work and heat, and verify
the fluctuation theorem for heat engines. We also derive the generating
function for the power and the efficiency and discuss their variances
for the Brownian CA engine.\renewcommand\theequation{S\arabic{equation}}

\section{Construction and optimization of a finite-time Brownian engine\label{sec:Construction-and-optimization}}

We attempt to construct and optimize a finite-time Brownian Carnot
engine with the microscopic model proposed in the main text. A Carnot
cycle consists of two isothermal processes and two adiabatic processes.
From Ref. \citep{salazar2019stochastic,Rubin1979a}, we know that
for a given duration of time and the initial and the final values
of the work parameter of an isothermal expansion (compression) process,
the protocol $\lambda(t)$ that maximizes the work output (minimizes
the work input) is an exponential function of time with additional
instantaneous jumps at the beginning and the end. The effective temperature
of the working substance is kept at a constant except for the sudden
jump processes. The jump-exponential-jump form of the optimal work
protocol $\lambda(t)$ and the constant effective temperature are
general properties of the optimal control, and are independent of
specific settings of the control time and the initial and final values
of the work parameter, as long as we maximize the work output.

We model the adiabatic process to be an instantaneous jump of the
work parameter $\lambda$, i.e., the process is so fast that no heat
is exchanged. Then, the equation of motion of the effective temperature
by Eq. (7) is reduced to 
\begin{equation}
\dot{\theta}(t)=\frac{\dot{\lambda}(t)}{\lambda(t)}\theta(t).\label{eq:adiabatic_theta}
\end{equation}
After doing integral on both sides of Eq. (\ref{eq:adiabatic_theta}),
we obtain the relation between the initial (final) value of the effective
temperature $\theta_{i}\,(\theta_{f})$ and the work parameter $\lambda_{i}\,(\lambda_{f})$
\begin{equation}
\frac{\theta_{i}}{\theta_{f}}=\frac{\lambda_{i}}{\lambda_{f}}.\label{eq:jump_temp}
\end{equation}
The work input in an adiabatic process is equal to the energy change

\begin{equation}
W=E_{f}-E_{i},\,\left\langle W\right\rangle =\frac{f_{n}}{2}(\theta_{f}-\theta_{i}),
\end{equation}
since there is no heat exchange.

To form a closed cycle, we need to connect the two isothermal processes
with the two adiabatic processes. The additional jumps and the adiabatic
processes (also instantaneous jumps) merge together. The control scheme
$\lambda(t)$ consists of two exponential protocols and two instantaneous
jump protocols (Fig. 1). It is natural to describe the cycle with
the work parameter at four ``corners'' in the $\lambda-\theta$
diagram (Fig. 1), and the duration $\tau_{H}\,(\tau_{C})$ of two
exponential protocols. Hereafter, we also use ``isothermal process''
to denote the exponential protocol, and ``adiabatic processes''
to denote the sudden jump protocol.

We formally write down the four processes as follows.
\global\long\def\theenumi{\Roman{enumi}}%
\global\long\def\labelenumi{(\theenumi)}%

\begin{enumerate}
\item \label{enu:Isothermal-compression.-Starting}Isothermal compression.
Starting from $\lambda_{1}$ at $t=0$, the control scheme of the
work parameter is $\lambda(t)=\lambda_{1}(\lambda_{2}/\lambda_{1})^{t/\tau_{C}}$,
where $\tau_{C}$ is the duration of the process. For such an exponential
protocol, the effective temperature of the working substance remains
at a constant 
\begin{equation}
\theta_{C}=\frac{\tau_{C}\Gamma_{C}}{\tau_{C}\Gamma_{C}-\ln r}T_{C}.\label{eq:workingfluidtemperature}
\end{equation}
with the quantity $r=\lambda_{2}/\lambda_{1}$ defined in Eq. (8)
of the main text. The heat released to the cold heat bath and the
work input in this process is

\begin{equation}
-\left\langle Q_{C}\right\rangle =\left\langle W_{1}\right\rangle =\frac{f_{n}}{2}\Gamma_{C}(\theta_{C}-T_{C})\tau_{C}.\label{eq:coldisothermal_heat}
\end{equation}

\item Adiabatic compression. The work parameter is quenched suddenly from
$\lambda_{2}$ to $\lambda_{3}$ with the effective temperature jumping
from $\theta_{C}$ to $\theta_{H}$ accordingly, satisfying $\theta_{H}/\theta_{C}=\lambda_{3}/\lambda_{2}$.
Work output in this adiabatic compression process is
\begin{equation}
-\left\langle W_{2}\right\rangle =-\frac{f_{n}}{2}(\theta_{H}-\theta_{C}).
\end{equation}
\item Isothermal expansion. Similar to process (\ref{enu:Isothermal-compression.-Starting}),
the work parameter is varied as $\lambda(t)=\lambda_{3}(\lambda_{4}/\lambda_{3})^{(t-\tau_{C})/\tau_{H}}$
with the duration $\tau_{H}$ of this process. The effective temperature
of the working substance, as a constant, is explicitly
\begin{equation}
\theta_{H}=\frac{\tau_{H}\Gamma_{H}}{\tau_{H}\Gamma_{H}+\ln r}T_{H}.\label{eq:workingfiulrtmeperature_hot}
\end{equation}
The heat absorbed from the hot bath and the work output in this process
is
\begin{equation}
\left\langle Q_{H}\right\rangle =-\left\langle W_{3}\right\rangle =\frac{f_{n}}{2}\Gamma_{H}(T_{H}-\theta_{H})\tau_{H}.\label{eq:hotisothermal_heat}
\end{equation}
\item Adiabatic expansion. Finally, the work parameter is quenched from
$\lambda_{4}$ to the initial value $\lambda_{1}$ abruptly. The values
of the work parameter satisfy $\theta_{H}/\theta_{C}=\lambda_{4}/\lambda_{1}$,
which ensures a closed cycle. The work output in this adiabatic process
is
\begin{equation}
-\left\langle W_{4}\right\rangle =-\frac{f_{n}}{2}(\theta_{C}-\theta_{H}).
\end{equation}
\end{enumerate}
The work output of a cycle is equal to
\begin{align}
-\left\langle W\right\rangle  & \coloneqq-\sum_{j=1}^{4}\left\langle W_{j}\right\rangle \nonumber \\
 & =\frac{f_{n}}{2}\left(\frac{\tau_{H}\Gamma_{H}\ln r}{\tau_{H}\Gamma_{H}+\ln r}T_{H}-\frac{\tau_{C}\Gamma_{C}\ln r}{\tau_{C}\Gamma_{C}-\ln r}T_{C}\right).
\end{align}
The average power and efficiency of this finite-time Brownian Carnot
engine are
\begin{equation}
\overline{P}\coloneqq\frac{-\left\langle W\right\rangle }{\tau_{H}+\tau_{C}}=\frac{f_{n}}{2(\tau_{H}+\tau_{C})}\left(\frac{\tau_{H}\Gamma_{H}\ln r}{\tau_{H}\Gamma_{H}+\ln r}T_{H}-\frac{\tau_{C}\Gamma_{C}\ln r}{\tau_{C}\Gamma_{C}-\ln r}T_{C}\right),\label{eq:power}
\end{equation}
and 
\begin{equation}
\eta\coloneqq-\frac{\left\langle W\right\rangle }{\left\langle Q_{H}\right\rangle }=1-\frac{1+(\tau_{H}\Gamma_{H})^{-1}\ln r}{1-(\tau_{C}\Gamma_{C})^{-1}\ln r}\frac{T_{C}}{T_{H}}.\label{eq:efficiency}
\end{equation}
To maximize the power, we choose a fixed $r$ and optimize $\overline{P}$
with respect to $\tau_{H}$ and $\tau_{C}$. By solving equations
$\partial\overline{P}/\partial\tau_{H}=0$ and $\partial\overline{P}/\partial\tau_{C}=0$,
we obtain the optimal duration of the two isothermal processes

\begin{align}
 & \tau_{H}^{\mathrm{max}}=\frac{\ln r\left(\sqrt{\Gamma_{H}T_{H}}+\sqrt{\Gamma_{C}T_{C}}\right)}{\Gamma_{H}\sqrt{\Gamma_{C}}\left(\sqrt{T_{H}}-\sqrt{T_{C}}\right)},\nonumber \\
 & \tau_{C}^{\mathrm{max}}=\frac{\ln r\left(\sqrt{\Gamma_{H}T_{H}}+\sqrt{\Gamma_{C}T_{C}}\right)}{\Gamma_{C}\sqrt{\Gamma_{H}}\left(\sqrt{T_{H}}-\sqrt{T_{C}}\right)}.\label{eq:max_time}
\end{align}

Substituting Eq. (\ref{eq:max_time}) into Eq. (\ref{eq:power}),
we find the maximum power with the fixed $r$ is
\begin{equation}
\overline{P}_{\mathrm{max}}=\frac{f_{n}\Gamma_{C}\Gamma_{H}\left(\sqrt{T_{H}}-\sqrt{T_{C}}\right)^{2}}{2\left(\sqrt{\Gamma_{H}}+\sqrt{\Gamma_{C}}\right)^{2}}.
\end{equation}
This result, however, is independent of $r$, and is the global maximum
of power. Similarly substituting Eq.(\ref{eq:max_time}) into Eq.
(\ref{eq:efficiency}), the efficiency at the maximum power is obtained
\begin{equation}
\eta_{\mathrm{EMP}}=1-\sqrt{\frac{T_{C}}{T_{H}}},
\end{equation}
 which is exactly the CA efficiency.

\section{the trade-off relation between power and efficiency\label{sec:tradeoffrelation}}

The trade-off relation between power and efficiency, namely the maximum
power under a given efficiency, has recently attracted much attention
in finite-time thermodynamics. Various trade-off relations have been
found under different circumstances \citep{shiraishi2016universal,Holubec2016,GonzalezAyala2017,pietzonka2018universal,Ma2018,Abiuso2020}.
Nevertheless, these relations either are obtained based on the low-dissipation
approximation, or are very loose constraints.

It is worth pointing out that in 1989, a tight trade-off relation
between power and efficiency has been obtained for the endoreversible
Carnot cycle with the assumption of Newton's cooling law \citep{Chen1989},
\begin{equation}
\overline{P}^{*}\propto\frac{\eta\left(\eta_{\mathrm{C}}-\eta\right)}{1-\eta}.\label{eq:tradeoff_yan-1}
\end{equation}
However, due to the lack of microscopic theory in their study, they
are unable to determine the control scheme associated with the maximum
power for a given efficiency.

Here we derive the control scheme $\lambda(t)$ associated with the
above tight trade-off relation between power and efficiency for the
finite-time Brownian Carnot engine. With the expressions of the power
and the efficiency, we introduce the function

\begin{equation}
L=\frac{f_{n}}{2(\tau_{H}+\tau_{C})}\left(\frac{\tau_{H}\Gamma_{H}\ln r}{\tau_{H}\Gamma_{H}+\ln r}T_{H}-\frac{\tau_{C}\Gamma_{C}\ln r}{\tau_{C}\Gamma_{C}-\ln r}T_{C}\right)+\mu\left(1-\frac{1+(\tau_{H}\Gamma_{H})^{-1}\ln r}{1-(\tau_{C}\Gamma_{C})^{-1}\ln r}\frac{T_{C}}{T_{H}}-\eta\right),
\end{equation}
where $\mu$ is the Lagrange multiplier to include the given efficiency
as the constraint. The maximum power for the given efficiency $\eta$
is obtained from 
\begin{align}
\frac{\partial L}{\partial\tau_{H}} & =0,\\
\frac{\partial L}{\partial\tau_{C}} & =0,\\
\frac{\partial L}{\partial\mu} & =0.
\end{align}
The solution to the above equations are

\begin{align}
\tau_{H}^{*} & =\frac{\ln r}{\sqrt{\Gamma_{H}}\left(\eta_{\mathrm{C}}-\eta\right)}\left(\frac{1-\eta}{\sqrt{\Gamma_{C}}}+\frac{1-\eta_{\mathrm{C}}}{\sqrt{\Gamma_{H}}}\right),\\
\tau_{C}^{*} & =\frac{\ln r}{\sqrt{\Gamma_{C}}\left(\eta_{\mathrm{C}}-\eta\right)}\left(\frac{1-\eta}{\sqrt{\Gamma_{C}}}+\frac{1-\eta_{\mathrm{C}}}{\sqrt{\Gamma_{H}}}\right),\\
\mu^{*} & =\frac{f_{n}\Gamma_{C}\Gamma_{H}T_{H}\left[(2-\eta)\eta-\eta_{C}\right]}{2(1-\eta)^{2}(\sqrt{\Gamma_{H}}+\sqrt{\Gamma_{C}})^{2}}.
\end{align}

Since we have adopted the exponential protocol, the control scheme
$\lambda(t)$ of the whole cycle for the CA engine is

\begin{equation}
\lambda(t)=\begin{cases}
\lambda_{1}(\lambda_{2}/\lambda_{1})^{t/\tau_{C}^{*}} & 0<t<\tau_{C}^{*}\\
\lambda_{3}(\lambda_{4}/\lambda_{3})^{(t-\tau_{C}^{*})/\tau_{H}^{*}} & \tau_{C}^{*}<t<\tau_{H}^{*}
\end{cases}.
\end{equation}
Substituting $\tau_{H}^{*}$ and $\tau_{C}^{*}$ into the expression
of the power, we obtain the maximum power for a given efficiency $\eta$
as

\begin{align}
\overline{P}^{*} & =\frac{f_{n}T_{H}\Gamma_{C}\Gamma_{H}}{2\left(\sqrt{\Gamma_{C}}+\sqrt{\Gamma_{H}}\right)^{2}}\frac{\eta(\eta_{C}-\eta)}{(1-\eta)},\label{eq:23}
\end{align}
which agrees with the trade-off relation obtained in Ref. \citep{Chen1989}.
Notice that Eq. (\ref{eq:23}) is independent of the compression ratio
$r$, and is also irrelevant to the symmetry ($\Gamma_{H}=\Gamma_{C}$)
of dissipation.

We emphasize that the dependence of the maximum power on the efficiency
is always in the factorized form, which is universal and independent
of the cooling rate $\Gamma_{H}\,(\Gamma_{C})$ in the two isothermal
processes. In comparison with Ref. \citep{shiraishi2016universal},
the trade-off relation (\ref{eq:23}) is tighter and this upper bound
of the power can be achieved with the explicit control scheme.

\section{Calculation of the generating function\label{sec:Calculation-of-the-1}}

Based on the method proposed in Ref. \citep{salazar2020work}, we
derive the joint generating function of work and heat for both the
finite-time driving process (finite-time isothermal process). The
system is coupled to only one heat bath with the exchanged heat $Q[X]$,
and the work parameter of the system is tuned with the performed work
$W[X]$. In stochastic thermodynamics, both heat and work is defined
on the trajectory $X=\{E(t)|0\leq t\leq\tau\}$ \citep{Sekimoto1998,salazar2020work}.
The joint generating function of work and heat is defined as the path
integral in the trajectory space

\begin{align}
I(u,s) & \coloneqq\left\langle e^{uQ+sW}\right\rangle \\
 & =\int D[X]p[X]\mathrm{e}^{uQ[X]+sW[X]},
\end{align}
where $p[X]$ denotes the probability of a given trajectory $X$.

The method \citep{salazar2020work} to be applied requires that
\begin{enumerate}
\item \label{enu:The-system-always}The system always obeys a Maxwell-Boltzmann
distribution in the energy space described by an effective temperature
$\theta$, whose evolution is governed by 
\begin{equation}
\dot{\theta}=\Phi_{t}(\theta),\label{eq:temp_evolution-1}
\end{equation}
Here, $\Phi_{t}(\theta)$ contains the contribution of the work. When
performed work performed on the system, the temperature of the system
increases. If no modulation is performed, $\Phi_{t}(\theta)=-\Gamma(\theta-T_{\mathrm{b}})$
is Newton's law of cooling with a constant cooling rate $\Gamma$.
$T_{\mathrm{b}}$ is the temperature of the bath.
\item \label{itema2-1}The increment in work is proportional to the stochastic
energy of the system,
\begin{equation}
dW=\alpha(t)E(t)dt,\label{eq:increment_work-1}
\end{equation}
where $\alpha_{t}$ is determined by the control protocol, and is
chosen as $\alpha_{t}=\dot{\lambda}/\lambda$ relating to the work
parameter $\lambda=\lambda(t)$; $E=E(t)$ is the stochastic energy
of the system affected by both the heat exchange and the work performed.
As a result, $\Phi_{t}(\theta)$ is explicitly 
\begin{equation}
\Phi_{t}(\theta)=-\Gamma(\theta-T_{\mathrm{b}})+\alpha_{t}\theta.
\end{equation}
\item The structure (shape) of the partition function does not depend on
$\lambda$. Generally, the partition function $Z_{\lambda}(\beta)$
should depend on the work parameter $\lambda$. The inverse temperature
is $\beta=1/\theta$, where we set $k_{\mathrm{B}}=1$ for convenience.
For example, for a quantum harmonic oscillator, the frequency determines
energy-level spacing which affects the $Z_{\lambda}(\beta)$. But
here, we require $Z_{\lambda}(\beta)$ to be in the same form of $\lambda$
which means
\begin{equation}
Z_{\lambda}(\beta)=g(\lambda)\times Z(\beta),
\end{equation}
where $g(\lambda)$ is a function of the work parameter, and $Z(\beta)$
characterize the form of the energy density.
\end{enumerate}
All conditions are satisfied in the current microscopic model. The
probability distribution of the energy is

\begin{equation}
p(E|\theta)=\frac{E^{\frac{f_{n}}{2}-1}\mathrm{e}^{-\beta E}}{Z(\beta)}.
\end{equation}
The partition function is $Z(\beta)=\int_{0}^{\infty}E^{f_{n}/2-1}\mathrm{e}^{-\beta E}dE=\beta^{-f_{n}/2}\Gamma(f_{n}/2)$\footnote{$\Gamma(x)=\int_{0}^{\infty}e^{-y}y^{x-1}dy$ is the Gamma function.},
which is also the normalization factor in the energy space. The average
energy is determined by
\begin{equation}
\left\langle E\right\rangle =-\frac{\partial\ln\left[Z(\beta)\right]}{\partial\beta}=\frac{f_{n}}{2}\theta.\label{eq:average_energy-1}
\end{equation}
Namely, the heat capacity of the working substance is $f_{n}/2$,
which remains a constant.

Along a given trajectory $X=\{E(t)|0\leq t\leq\tau\}$, the increment
of work at time $t$ is

\begin{equation}
dW=\alpha_{t}Edt,
\end{equation}
and the increment of heat is obtained inversely from the first law

\begin{equation}
dQ=dE-\alpha_{t}Edt.
\end{equation}

\subsection{Discretization to calculate the joint generating function of work
and heat}

We discretize the dynamics with $t_{j}=j\epsilon$, $j=0,1,...,N,N+1$
and $\epsilon=\tau/(N+1)$, the thermal distribution of the system
is $p(E_{j}|\theta_{j})$ with the inverse temperature $\beta_{j}=1/\theta_{j}$
at $t_{j}$. The work parameter $\lambda(t)$ remains a constant during
each time slice $t_{j}<t<t_{j+1}$, and is quenched at the moment
$t_{j}$. As a result, the work $W_{j}=\ln[\lambda(t_{j}^{+})/\lambda(t_{j}^{-})]E_{j}\approx\alpha_{j}E_{j}\epsilon$
is performed at each moment $t_{j}$ with the neglected exchanged
heat, while the heat $Q_{j}=E_{j+1}-E_{j}-\alpha_{j}E_{j}\epsilon$
is generated during each time slice. The evolution between every adjacent
moment $t_{j}$ and $t_{j+1}$ can be described by the transition
probability $\mathcal{R}_{j+1}^{j}=\mathcal{R}(E_{j+1},t_{j+1};E_{j},t_{j})$
(also named as the propagator). To ensure the condition (\ref{enu:The-system-always}),
the propagator $\mathcal{R}_{j+1}^{j}$ should maps a local equilibrium
state to another local equilibrium state

\begin{equation}
\int dE_{j}p(E_{j}|\theta_{j})\mathcal{R}_{j+1}^{j}=p(E_{j+1}|\theta_{j+1}).
\end{equation}
The temperature of the next step is 
\begin{equation}
\theta_{j+1}=\theta_{j}+\Phi_{j}(\theta_{j})\epsilon,\label{eq:thetaj+1}
\end{equation}
where we denote $\Phi_{j}(\cdot)=\Phi_{t_{j}}(\cdot)$ for simplicity

We calculate the joint generating function $I(u,s)$ as follows. We
rewrite the joint generating function with the $N+1$ discrete steps

\begin{equation}
I(u,s)=\int\prod_{j=0}^{N+1}dE_{j}\prod_{j=0}^{N}\left(\mathcal{R}_{j+1}^{j}\right)p(E_{0}|\tilde{\psi}_{0})\prod_{j=0}^{N}e^{uQ_{j}+sW_{j}},
\end{equation}
where the initial value $\tilde{\psi}_{0}$ is a temperature-like
quantity. We use a different notation $\psi$ to distinguish with
the temperature $\theta$. The meaning of ``tilde'' will be clarified
later {[}by Eq. (\ref{eq:psi_tilde}){]}. If there is no previous
process, $\tilde{\psi}_{0}=\theta_{0}$ is the local temperature of
the system. Otherwise, the initial value $\tilde{\psi}_{0}$ is obtained
from the previous process.

From the first law, we can rewrite

\begin{align}
uQ_{j}+sW_{j} & =(s-u)\alpha_{j}E_{j}\epsilon+u(E_{j+1}-E_{j}).
\end{align}
We then consider the first step

\begin{align}
 & \mathcal{R}_{1}^{0}p(E_{0}|\tilde{\psi}_{0})e^{(s-u)W_{0}+u(E_{1}-E_{0})}\\
= & \mathcal{R}_{1}^{0}\frac{e^{-E_{0}/\tilde{\psi}_{0}}}{Z(1/\tilde{\psi}_{0})}e^{(s-u)\epsilon\alpha_{0}E_{0}+u(E_{1}-E_{0})}\\
= & \mathcal{R}_{1}^{0}\frac{e^{-E_{0}/\psi_{0}}}{Z(1/\tilde{\psi}_{0})}e^{uE_{1}},
\end{align}
where $\psi_{0}$ is 
\begin{equation}
\psi_{0}=1/[1/\tilde{\psi}_{0}-(s-u)\epsilon\alpha_{0}+u].\label{eq:determining_psi_0}
\end{equation}
 After doing the integral over $E_{0}$, we obtain

\begin{align}
 & e^{uE_{1}}\int dE_{0}\mathcal{R}_{1}^{0}\frac{e^{-E_{0}/\psi_{0}}}{Z(1/\tilde{\psi}_{0})}\\
= & e^{uE_{1}}\frac{Z(1/\psi_{0})}{Z(1/\tilde{\psi}_{0})}\int dE_{0}\mathcal{R}_{1}^{0}\frac{e^{-E_{0}/\psi_{0}}}{Z(1/\psi_{0})}\\
= & e^{uE_{1}}\frac{Z(1/\psi_{0})}{Z(1/\tilde{\psi}_{0})}\frac{e^{-E_{1}/\varphi_{1}}}{Z(1/\varphi_{1})},
\end{align}
where the intermediate variable $\varphi_{1}$, according to Eq. (\ref{eq:thetaj+1}),
is obtained as

\begin{align}
\varphi_{1} & =\psi_{0}+\Phi_{0}(\psi_{0})\epsilon\\
 & =\psi_{0}-\Gamma(\psi_{0}-T_{\mathrm{b}})\epsilon+\alpha_{0}\psi_{0}\epsilon.
\end{align}

We next calculate the second step with the integral over $E_{1}$

\begin{align}
 & \frac{Z(1/\psi_{0})}{Z(1/\tilde{\psi}_{0})}\int dE_{1}\mathcal{R}_{2}^{1}\frac{e^{-E_{1}/\varphi_{1}}e^{uE_{1}}}{Z(1/\varphi_{1})}e^{(s-u)W_{1}+u(E_{2}-E_{1})}\\
= & \frac{Z(1/\psi_{0})}{Z(1/\tilde{\psi}_{0})}e^{uE_{2}}\int dE_{1}\mathcal{R}_{2}^{1}\frac{e^{-\overset{\coloneqq1/\psi_{1}}{\overbrace{[1/\varphi_{1}-(s-u)\alpha_{1}\epsilon]}}E_{1}}}{Z(1/\varphi_{1})}\label{eq:secondlineofsecondstep}\\
= & \frac{Z(1/\psi_{0})}{Z(1/\tilde{\psi}_{0})}\frac{Z(1/\psi_{1})}{Z(1/\varphi_{1})}\frac{e^{uE_{2}}e^{-E_{2}/\varphi_{2}}}{Z(1/\varphi_{2})}.
\end{align}
In Eq. (\ref{eq:secondlineofsecondstep}), we have set 
\begin{equation}
\frac{1}{\psi_{1}}=\frac{1}{\varphi_{1}}-(s-u)\alpha_{1}\epsilon.
\end{equation}
It is similarly to do the integral over $E_{2},E_{3},...E_{N}$ and
the result is

\begin{align}
 & \int\prod_{j=0}^{N}dE_{j}\prod_{j=0}^{N}\left(\mathcal{R}_{j+1}^{j}\right)p(E_{0}|\tilde{\psi}_{0})\prod_{j=0}^{N}e^{uQ_{j}+sW_{j}}\\
= & \frac{Z(1/\psi_{0})}{Z(1/\tilde{\psi}_{0})}\frac{Z(1/\psi_{1})}{Z(1/\varphi_{1})}...\frac{Z(1/\psi_{N})}{Z(1/\varphi_{N})}\frac{e^{-(1/\varphi_{N+1}-u)E_{N+1}}}{Z(1/\varphi_{N+1})}.
\end{align}
We integrate over $E_{N+1}$ and obtain

\begin{equation}
I(u,s)=\frac{Z(1/\psi_{0})}{Z(1/\tilde{\psi}_{0})}\frac{Z(1/\psi_{1})}{Z(1/\varphi_{1})}...\frac{Z(1/\psi_{N})}{Z(1/\varphi_{N})}\frac{Z(1/\tilde{\psi}_{N+1})}{Z(1/\varphi_{N+1})},\label{eq:generating_function_discreteversion}
\end{equation}
where

\begin{equation}
\tilde{\psi}_{N+1}=\frac{1}{1/\varphi_{N+1}-u}.\label{eq:psitilde}
\end{equation}
The reduction formulas are

\begin{align}
\varphi_{n} & =\psi_{n-1}+\Phi_{n-1}(\psi_{n-1})\epsilon\\
\psi_{n} & =\frac{1}{1/\varphi_{n}-(s-u)\alpha_{n}\epsilon},
\end{align}
which lead to

\begin{equation}
\frac{\psi_{n}-\psi_{n-1}}{\epsilon}=\Phi_{n-1}(\psi_{n-1})+(s-u)\alpha_{n}\psi_{n-1}^{2}.
\end{equation}

The continuum limit is

\begin{equation}
\frac{d\psi}{dt}=\Phi_{t}(\psi)+(s-u)\alpha_{t}\psi^{2}.\label{eq:continuumlimitdifferential_equation}
\end{equation}
The initial condition is obtained from Eq. (\ref{eq:determining_psi_0})
as

\begin{equation}
\psi(0)=\frac{\tilde{\psi}_{0}}{1+u\tilde{\psi}_{0}}.\label{eq:initial_condition_psi}
\end{equation}
At the end $t=\tau$, Eq. (\ref{eq:psitilde}) leads to the final
condition

\begin{equation}
\tilde{\psi}(\tau)=\frac{\psi(\tau)}{1-u\psi(\tau)}.\label{eq:final_condition_psi}
\end{equation}
We can define a shifted temperature-like variable $\tilde{\psi}(t)$
as

\begin{equation}
\tilde{\psi}(t)\coloneqq\frac{\psi(t)}{1-u\psi(t)}.\label{eq:psi_tilde}
\end{equation}
The corresponding differential equation is
\begin{equation}
\frac{d\tilde{\psi}}{dt}=(1+u\tilde{\psi})^{2}\Phi_{t}\left(\frac{\tilde{\psi}}{1+u\tilde{\psi}}\right)+(s-u)\alpha_{t}\tilde{\psi}^{2},\label{eq:psitildedifferential_equation}
\end{equation}
and the initial value is $\tilde{\psi}(0)=\tilde{\psi}_{0}$.

The joint generating function Eq. (\ref{eq:generating_function_discreteversion})
becomes

\begin{align}
\ln I(u,s) & =\ln\frac{Z(1/\psi_{0})}{Z(1/\tilde{\psi}_{0})}+\sum_{n=1}^{N}\ln\frac{Z(1/\psi_{n})}{Z(1/\varphi_{n})}+\ln\frac{Z(1/\tilde{\psi}_{N+1})}{Z(1/\varphi_{N+1})}.
\end{align}
The first and the third terms are explicitly

\begin{align}
\ln\frac{Z(1/\psi_{0})}{Z(1/\tilde{\psi}_{0})} & \approx\ln\frac{Z(1/\tilde{\psi}_{0}+u)}{Z(1/\tilde{\psi}_{0})}\\
 & =-\frac{f_{n}}{2}\ln\left(1+u\tilde{\psi}_{0}\right),
\end{align}
and

\begin{align}
\ln\frac{Z(1/\tilde{\psi}_{N+1})}{Z(1/\varphi_{N+1})} & \approx\ln\frac{Z(1/\psi_{N+1}-u)}{Z(1/\psi_{N+1})}\\
 & =-\frac{f_{n}}{2}\ln\left(1-u\psi(\tau)\right)\\
 & =\frac{f_{n}}{2}\ln\left(1+u\tilde{\psi}(\tau)\right).
\end{align}
The second term is explicitly

\begin{align}
\sum_{n=1}^{N}\ln\frac{Z(1/\psi_{n})}{Z(1/\varphi_{n})} & =\sum_{n=1}^{N}\ln\frac{Z(1/\psi_{n})}{Z(1/\psi_{n}+(s-u)\alpha_{n}\epsilon)}\\
 & \approx-\sum_{n=1}^{N}(s-u)\alpha_{n}\frac{Z^{\prime}(1/\psi_{n})}{Z(1/\psi_{n})}\epsilon\\
 & =-\int_{0}^{\tau}(s-u)\alpha_{t}\frac{Z^{\prime}(1/\psi(t))}{Z(1/\psi(t))}dt\\
 & =(s-u)\int_{0}^{\tau}\alpha_{t}\left\langle E\right\rangle (\psi(t))dt,
\end{align}
where the average internal energy is

\begin{equation}
\left\langle E\right\rangle (\theta)=-\frac{\partial\ln Z}{\partial\beta}\big|_{\beta=1/\theta}=\frac{f_{n}}{2}\theta.
\end{equation}
The final result of the joint generating function is

\begin{align}
\ln I(u,s) & =\frac{f_{n}}{2}\left[\ln\left(\frac{1+u\tilde{\psi}(\tau)}{1+u\tilde{\psi}_{0}}\right)+(s-u)\int_{0}^{\tau}\alpha_{t}\psi(t)dt\right].
\end{align}
Notice that the temperature-like variable $\psi(t)$ is solved from
the differential equation (\ref{eq:continuumlimitdifferential_equation})
with the initial condition (\ref{eq:initial_condition_psi}) and the
final condition (\ref{eq:final_condition_psi}).

As follows, we discuss the joint generating function for the sudden
jump process and the exponential protocol process. These two processes
are used to form the finite-time Carnot cycle.

\subsection{Sudden jump process}

In a sudden jump process, work is performed with the neglected heat
exchange. The work parameter is quenched from $\lambda^{-}$ to $\lambda^{+}$
with temperature-like quantity changed from $\tilde{\psi}^{-}$ to
$\tilde{\psi}^{+}$. No exchanged heat is produced in such a sudden
process, and we can set $u=0$. Therefore, $\tilde{\psi}$ becomes
the same as $\psi$ in this process. The differential equation (\ref{eq:psitildedifferential_equation})
{[}or Eq. (\ref{eq:continuumlimitdifferential_equation}){]} becomes

\begin{equation}
\frac{d\tilde{\psi}}{dt}=\alpha_{t}\tilde{\psi}+s\alpha_{t}\tilde{\psi}^{2},\label{eq:psitildedifferential_equation-1}
\end{equation}
The change of the energy comes from the work 
\begin{align}
d\tilde{\psi} & =\tilde{\psi}\left(1+s\tilde{\psi}\right)\frac{d\lambda}{\lambda},
\end{align}
and $\tilde{\psi}$ becomes a function of $\lambda$ in this process,
explicitly as

\begin{align}
\frac{\tilde{\psi}}{1+\tilde{\psi}s} & =\frac{\lambda}{\lambda^{-}}\frac{\tilde{\psi}^{-}}{1+\tilde{\psi}^{-}s}.
\end{align}
Thus, the final value $\tilde{\psi}^{+}$ is associated with $\tilde{\psi}^{-}$
as

\begin{equation}
\frac{\tilde{\psi}^{+}}{1+\tilde{\psi}^{+}s}=\frac{\lambda^{+}}{\lambda^{-}}\frac{\tilde{\psi}^{-}}{1+\tilde{\psi}^{-}s}.\label{eq:psi+associated_in_adiabatic_process}
\end{equation}
Notice that $\tilde{\psi}^{+}$ serves as the initial value $\tilde{\psi}_{0}$
of the next process. If there is a previous process, $\tilde{\psi}^{-}$
is substituted with $\tilde{\psi}(\tau)$ of the previous process.
The joint generating function of the sudden jump process is

\begin{equation}
I_{\mathrm{jump}}(u,s)=\left[1+\left(1-\frac{\lambda^{+}}{\lambda^{-}}\right)\tilde{\psi}^{-}s\right]^{-\frac{f_{n}}{2}}.\label{eq:i_JUMP-PROCESS}
\end{equation}

\subsection{Exponential protocol process}

Equation (\ref{eq:continuumlimitdifferential_equation}) becomes a
Ricatti equation (time-independent) for a exponential process $\alpha_{t}=\alpha=\mathrm{const}$,
and is explicitly

\begin{equation}
\dot{\psi}=-\Gamma(\psi-T_{\mathrm{b}})+\alpha\psi+(s-u)\alpha\psi^{2}.\label{eq:95}
\end{equation}
We solve the above differential equation with the initial condition
$\psi(0)=\tilde{\psi}_{0}/(1+u\tilde{\psi}_{0})$. The solution to
Eq. (\ref{eq:95}) is 
\begin{align}
\psi(t) & =\frac{\Gamma-\alpha+\sqrt{D(s-u)}\tan\left(\frac{c_{1}+t}{2}\sqrt{D(s-u)}\right)}{2\alpha(s-u)},
\end{align}
where

\begin{equation}
D(x)=4\alpha\Gamma T_{\mathrm{b}}x-(\Gamma-\alpha)^{2}.
\end{equation}
The joint generating function of the process is

\begin{equation}
\ln I_{\mathrm{exp}}(u,s)=\frac{f_{n}}{2}\left[\ln\left(\frac{1+u\tilde{\psi}(\tau)}{1+u\tilde{\psi}_{0}}\right)+(s-u)\alpha\int_{0}^{\tau}\psi(t)dt\right].\label{eq:lnI_exp}
\end{equation}
The initial condition gives

\begin{align}
\psi(0) & =\frac{\Gamma-\alpha+\sqrt{D(s-u)}\tan\left(\frac{c_{1}}{2}\sqrt{D(s-u)}\right)}{2\alpha(s-u)},
\end{align}
which determine the constant $c_{1}$ as

\begin{equation}
\tan\left(\frac{c_{1}}{2}\sqrt{D(s-u)}\right)=\frac{2\alpha(s-u)\psi(0)-(\Gamma-\alpha)}{\sqrt{D(s-u)}}.
\end{equation}
The integral in Eq. (\ref{eq:lnI_exp}) is rewritten as

\begin{align}
(s-u)\alpha\int_{0}^{\tau}\psi(t)dt & =\frac{\Gamma-\alpha}{2}\tau+\ln\left[\frac{\cos\left(\frac{c_{1}}{2}\sqrt{D(s-u)}\right)}{\cos\left(\frac{c_{1}+t}{2}\sqrt{D(s-u)}\right)}\right]\\
 & =\frac{\Gamma-\alpha}{2}\tau-\ln\left[\cos\left(\frac{t}{2}\sqrt{D(s-u)}\right)-\frac{2\alpha(s-u)\psi(0)-(\Gamma-\alpha)}{\sqrt{D(s-u)}}\sin\left(\frac{t}{2}\sqrt{D(s-u)}\right)\right].
\end{align}
We obtain the joint generating function as 
\begin{equation}
I_{\mathrm{exp}}(u,s)=\left\{ \frac{\exp\left[(\Gamma-\alpha)\tau/2\right][1+u\tilde{\psi}(\tau)]/[1+u\tilde{\psi}_{0}]}{\cos\left(\frac{t}{2}\sqrt{D(s-u)}\right)-\frac{2\alpha(s-u)\psi(0)-(\Gamma-\alpha)}{\sqrt{D(s-u)}}\sin\left(\frac{t}{2}\sqrt{D(s-u)}\right)}\right\} ^{\frac{f_{n}}{2}}.\label{eq:I_exp}
\end{equation}
If there is a previous process, we need to substitute $\tilde{\psi}_{0}$
as the final value $\tilde{\psi}(\tau)$ of the previous process.

\subsection{Joint generating function of the finite-time Carnot cycle}

Now we calculate the joint generating function of the finite-time
Carnot cycle. The heat is absorbed from the hot bath and released
to the cold, namely, $Q_{H}>0$ and $Q_{C}<0$. The joint generating
function $I(u_{H},u_{C},s)$ of work and heat for a whole finite-time
cycle is

\begin{equation}
I(u_{H},u_{C},s)=\int D[X]p[X]\mathrm{e}^{uQ_{H}[X]+u_{C}Q_{C}[X]+sW[X]}.
\end{equation}
Notice that the system is in contact with only one heat bath at a
time. We start from A point (Fig. 1 in main text) and calculate the
joint generating function of a whole cycle.

\subsubsection{Process $\mathrm{I}$ Isothermal compression}

We define the compression ratio as $r=\lambda_{2}/\lambda_{1}$. In
this process, the work parameter is varied exponentially with the
time

\begin{equation}
\lambda(t)=\lambda_{1}r^{\frac{t}{\tau_{C}}}.
\end{equation}
The quantity $\alpha$ of this process is explicitly $\alpha_{C}=\ln r/\tau_{C}$.
The initial temperature of the working substance is (which is also
the initial condition)
\begin{align}
\tilde{\psi}_{0} & =\theta_{C}=\frac{\Gamma_{C}}{\Gamma_{C}-\alpha_{C}}T_{C}.
\end{align}
If we consider the heat engine runs for many cycles, $\tilde{\psi}_{0}$
takes the final value of $\tilde{\psi}(\tau_{\mathrm{t}})$ of the
previous cycle. According to Eq. (\ref{eq:I_exp}), the joint generating
function of this process is

\begin{equation}
I_{\mathrm{I}}(u_{H},u_{C},s)=\left\{ \frac{\exp\left[(\Gamma_{C}-\alpha_{C})\tau_{C}/2\right][1+u_{C}\tilde{\psi}(\tau_{C}^{-})]/[1+u_{C}\tilde{\psi}_{0}]}{\cos\left(\Omega_{C}\tau_{C}\right)-\frac{\alpha_{C}(s-u_{C})\psi(0)-(\Gamma_{C}-\alpha_{C})/2}{\Omega_{C}}\sin\left(\Omega_{C}\tau_{C}\right)}\right\} ^{\frac{f_{n}}{2}},
\end{equation}
where the initial value is $\psi(0)=\tilde{\psi}_{0}/(1+u_{C}\tilde{\psi}_{0})$,
$\tau_{C}^{-}$ represents the moment of the end of process $\mathrm{I}$,
and the frequency $\Omega_{C}$ is

\begin{align}
\Omega_{C} & =\frac{\sqrt{D_{C}(s-u_{C})}}{2}=\sqrt{\alpha_{C}\Gamma_{C}T_{C}(s-u_{C})-(\Gamma_{C}-\alpha_{C})^{2}/4}.
\end{align}
Notice that the expression $I_{\mathrm{I}}$ does not contain $u_{H}$.
But if there exist a previous cycle, $\tilde{\psi}_{0}$ also relies
on $u_{H}$. Therefore, we always write $I(u_{H},u_{C},s)$ for all
the processes in a finite-time heat engine cycle.

\subsubsection{Process $\mathrm{II}$ Adiabatic compression.}

At the end of process $\mathrm{I}$, the temperature-like quantity
is $\tilde{\psi}(\tau_{C})$, which is plugged as the initial value
for the process $\mathrm{II}$ as a sudden quench of the work parameter
$\lambda$ from $\lambda_{2}$ to $\lambda_{3}$ at the moment $t=\tau_{C}$.
Notice that the ratio of the work parameter in this process satisfy
$\lambda_{3}/\lambda_{2}=\theta_{H}/\theta_{C}$. According to Eq.
(\ref{eq:psi+associated_in_adiabatic_process}), the temperature-like
quantity after the quench $\tilde{\psi}_{\tau_{C}}^{+}$ is associated
with $\tilde{\psi}_{\tau_{C}}^{-}=\tilde{\psi}(\tau_{C}^{-})$ as

\begin{align}
\tilde{\psi}_{\tau_{C}}^{+} & =\frac{\lambda_{3}\tilde{\psi}(\tau_{C}^{-})}{\lambda_{2}+s(\lambda_{2}-\lambda_{3})\tilde{\psi}(\tau_{C}^{-})}\label{eq:psi+associated_in_adiabatic_process-1}\\
 & =\frac{\tilde{\psi}(\tau_{C}^{-})}{1-\eta-s\eta\tilde{\psi}(\tau_{C}^{-})},\label{eq:psi_tauC^+}
\end{align}
where the efficiency of the finite-time Carnot cycle is $\eta=1-\theta_{C}/\theta_{H}$.
According to Eq. (\ref{eq:i_JUMP-PROCESS}), the joint generating
function of process $\mathrm{II}$ is

\begin{equation}
I_{\mathrm{II}}(u_{H},u_{C},s)=\left[1-\frac{s\eta}{1-\eta}\tilde{\psi}(\tau_{C}^{-})\right]^{-\frac{f_{n}}{2}}.
\end{equation}
Notice that the value of the temperature-like quantity $\tilde{\psi}(\tau_{C}^{-})$
depends on $u_{C}$ and $s$.

\subsubsection{Process $\mathrm{III}$ Isothermal expansion}

The work parameter is varied exponentially with the time

\begin{equation}
\lambda(t)=\lambda_{3}r^{-\frac{t-\tau_{C}}{\tau_{H}}}.
\end{equation}
The quantity $\alpha$ of this process is explicitly $\alpha_{H}=-\ln r/\tau_{H}$.
The initial condition of this process is

\begin{equation}
\tilde{\psi}(\tau_{C}^{+})=\tilde{\psi}_{\tau_{C}}^{+},
\end{equation}
or

\begin{equation}
\psi(\tau_{C}^{+})=\frac{\tilde{\psi}_{\tau_{C}}^{+}}{1+u_{H}\tilde{\psi}_{\tau_{C}}^{+}}.
\end{equation}
The adiabatic process is completed suddenly at the moment $t=\tau_{C}$,
and we use $\tau_{C}^{+}$ to indicate the beginning of process $\mathrm{III}$.
Equation (\ref{eq:I_exp}) gives the joint generating function of
this process 
\begin{equation}
I_{\mathrm{III}}(u_{H},u_{C},s)=\left\{ \frac{\exp\left[(\Gamma_{H}-\alpha_{H})\tau_{H}/2\right]\frac{1+u_{H}\tilde{\psi}(\tau_{C}+\tau_{H}^{-})}{1+u_{H}\tilde{\psi}_{\tau_{C}}^{+}}}{\cos\left(\Omega_{H}\tau_{H}\right)-\frac{\alpha_{H}(s-u_{H})\psi(\tau_{C}^{+})-(\Gamma_{H}-\alpha_{H})/2}{\Omega_{H}}\sin\left(\Omega_{H}\tau_{H}\right)}\right\} ^{\frac{f_{n}}{2}},\label{eq:I_exp-1}
\end{equation}
where $\tau_{C}+\tau_{H}^{-}$ represents the moment of the end of
process $\mathrm{III}$, and the frequency $\Omega_{H}$ is

\begin{align}
\Omega_{H} & =\frac{1}{2}\sqrt{D_{H}(s-u_{H})}=\sqrt{\alpha_{H}\Gamma_{H}T_{H}(s-u_{H})-(\Gamma_{H}-\alpha_{H})^{2}/4}.
\end{align}

\subsubsection{Process $\mathrm{IV}$ Adiabatic expansion}

The last adiabatic is a sudden quench with the work parameter $\lambda$
changing from $\lambda_{4}$ to $\lambda_{1}$ at the moment $t=\tau_{C}+\tau_{H}$.
The ratio of the work parameter satisfies $\lambda_{1}/\lambda_{4}=\theta_{C}/\theta_{H}$.
According to Eq. (\ref{eq:psi+associated_in_adiabatic_process}),
the temperature-like quantity after the quench $\tilde{\psi}_{\tau_{C}+\tau_{H}}^{+}$
is associated with $\tilde{\psi}_{\tau_{C}+\tau_{H}}^{-}=\tilde{\psi}(\tau_{C}+\tau_{H}^{-})$
as

\begin{align}
\tilde{\psi}_{\tau_{C}+\tau_{H}}^{+} & =\frac{\lambda_{1}\tilde{\psi}(\tau_{C}+\tau_{H}^{-})}{\lambda_{4}+s(\lambda_{4}-\lambda_{1})\tilde{\psi}(\tau_{C}+\tau_{H}^{-})}\label{eq:psi+associated_in_adiabatic_process-1-1}\\
 & =\frac{(1-\eta)\tilde{\psi}(\tau_{C}+\tau_{H}^{-})}{1+s\eta\tilde{\psi}(\tau_{C}+\tau_{H}^{-})},
\end{align}
According to Eq. (\ref{eq:i_JUMP-PROCESS}), the joint generating
function of process $\mathrm{IV}$ is

\begin{equation}
I_{\mathrm{IV}}(u_{H},u_{C},s)=\left[1+s\eta\tilde{\psi}(\tau_{C}+\tau_{H}^{-})\right]^{-\frac{f_{n}}{2}}.\label{eq:i_JUMP-PROCESS-1}
\end{equation}

\subsubsection{Joint generating function of a whole cycle}

The joint generating function $I_{\mathrm{cycle}}(u_{H},u_{C},s)$
of a whole cycle is

\begin{align}
I_{\mathrm{cycle}} & =I_{\mathrm{I}}\times I_{\mathrm{II}}\times I_{\mathrm{III}}\times I_{\mathrm{IV}}\\
 & =\left\{ \frac{e^{(\Gamma_{C}\tau_{C}+\Gamma_{H}\tau_{H})/2}\left[1-\frac{s\eta}{1-\eta}\tilde{\psi}(\tau_{C}^{-})\right]^{-1}\left[1+s\eta\tilde{\psi}(\tau_{C}+\tau_{H}^{-})\right]^{-1}\frac{1+u_{C}\tilde{\psi}(\tau_{C}^{-})}{1+u_{C}\tilde{\psi}_{0}}\frac{1+u_{H}\tilde{\psi}(\tau_{C}+\tau_{H}^{-})}{1+u_{H}\tilde{\psi}_{\tau_{C}}^{+}}}{\left[\cos\left(\Omega_{C}\tau_{C}\right)-\frac{\alpha_{C}(s-u_{C})\psi(0)-(\Gamma_{C}-\alpha_{C})/2}{\Omega_{C}}\sin\left(\Omega_{C}\tau_{C}\right)\right]\left[\cos\left(\Omega_{H}\tau_{H}\right)-\frac{\alpha_{H}(s-u_{H})\psi(\tau_{C}^{+})-(\Gamma_{H}-\alpha_{H})/2}{\Omega_{H}}\sin\left(\Omega_{H}\tau_{H}\right)\right]}\right\} ^{\frac{f_{n}}{2}},\label{eq:I_cycle_uHuCs}
\end{align}
The evolution of $\psi(t)$ is governed by

\begin{align}
\dot{\psi} & =\begin{cases}
-\Gamma_{C}(\psi-T_{C})+\alpha_{C}\psi+(s-u_{C})\alpha_{C}\psi^{2}, & 0<t<\tau_{C},\\
-\Gamma_{H}(\psi-T_{H})+\alpha_{H}\psi+(s-u_{H})\alpha_{H}\psi^{2}, & \tau_{C}<t<\tau_{H}.
\end{cases}\label{eq:95-1}
\end{align}
The initial $\psi(0)$ and the connecting conditions $\psi(\tau_{C}^{+})$
are

\begin{align}
\psi(0) & =\frac{\tilde{\psi}_{0}}{1+u_{C}\tilde{\psi}_{0}}\\
\psi(\tau_{C}^{+}) & =\frac{\tilde{\psi}_{\tau_{C}}^{+}}{1+u_{H}\tilde{\psi}_{\tau_{C}}^{+}}.
\end{align}
where $\tilde{\psi}_{\tau_{C}}^{+}$ is associated with $\tilde{\psi}(\tau_{C}^{-})$
through process $\mathrm{II}$. As follows, we give the generating
functions for work and efficiency respectively.

\subsubsection{Generating function for work}

By setting $u_{C}=0$ and $u_{H}=0$, the generating function for
work is $I_{\mathrm{cycle}}^{W}(s)=I_{\mathrm{cycle}}(0,0,s),$ explicitly
as

\begin{align}
I_{\mathrm{cycle}}^{W}(s) & =\left\{ \frac{e^{(\Gamma_{C}\tau_{C}+\Gamma_{H}\tau_{H})/2}\left[1-\frac{s\eta}{1-\eta}\psi(\tau_{C}^{-})\right]^{-1}\left[1+s\eta\psi(\tau_{C}+\tau_{H}^{-})\right]^{-1}}{\left[\cos\left(\Omega_{C}\tau_{C}\right)-\frac{\alpha_{C}s\psi(0)-(\Gamma_{C}-\alpha_{C})/2}{\Omega_{C}}\sin\left(\Omega_{C}\tau_{C}\right)\right]\left[\cos\left(\Omega_{H}\tau_{H}\right)-\frac{\alpha_{H}s\psi(\tau_{C}^{+})-(\Gamma_{H}-\alpha_{H})/2}{\Omega_{H}}\sin\left(\Omega_{H}\tau_{H}\right)\right]}\right\} ^{\frac{f_{n}}{2}}
\end{align}
with $\Omega_{C}=\sqrt{\alpha_{C}\Gamma_{C}T_{C}s-(\Gamma_{C}-\alpha_{C})^{2}/4}$
and $\Omega_{H}=\sqrt{\alpha_{H}\Gamma_{H}T_{H}s-(\Gamma_{H}-\alpha_{H})^{2}/4}.$
Notice that Eq. (\ref{eq:psi_tilde}) indicates $\tilde{\psi}(t)=\psi(t)$
in this situation.

\subsubsection{Generating function for efficiency}

By setting $u_{H}=s\eta$ and $u_{C}=0$, the generating function
for efficiency is $I_{\mathrm{cycle}}^{\zeta}(s)=I_{\mathrm{cycle}}(s\eta,0,s)$,
explicitly as

\begin{equation}
I_{\mathrm{cycle}}^{\zeta}(s)=\left\{ \frac{e^{(\Gamma_{C}\tau_{C}+\Gamma_{H}\tau_{H})/2}\left[1-\frac{s\eta}{1-\eta}\tilde{\psi}_{\tau_{C}}^{-}\right]^{-1}\left[1+s\eta\tilde{\psi}_{\tau_{C}}^{+}\right]^{-1}}{\left[\cos\left(\Omega_{C}\tau_{C}\right)-\frac{\alpha_{C}s\psi(0)-(\Gamma_{C}-\alpha_{C})/2}{\Omega_{C}}\sin\left(\Omega_{C}\tau_{C}\right)\right]\left[\cos\left(\Omega_{H}\tau_{H}\right)-\frac{\alpha_{H}(s-s\eta)\psi(\tau_{C}^{+})-(\Gamma_{H}-\alpha_{H})/2}{\Omega_{H}}\sin\left(\Omega_{H}\tau_{H}\right)\right]}\right\} ^{\frac{f_{n}}{2}},\label{eq:I_cycle^zeta}
\end{equation}
with $\Omega_{C}=\sqrt{\alpha_{C}\Gamma_{C}T_{C}s-(\Gamma_{C}-\alpha_{C})^{2}/4}$
and $\Omega_{H}=\sqrt{\alpha_{H}\Gamma_{H}T_{H}(s-u_{H})-(\Gamma_{H}-\alpha_{H})^{2}/4}.$.
Since $u_{C}=0$, $\tilde{\psi}(t)=\psi(t)$ during process $\mathrm{I}$
($0<t<\tau_{C}$). Plugging Eq. (\ref{eq:psi_tauC^+}) into Eq. (\ref{eq:I_cycle^zeta}),
we obtain

\begin{equation}
I_{\mathrm{cycle}}^{\zeta}(s)=\left\{ \frac{e^{(\Gamma_{C}\tau_{C}+\Gamma_{H}\tau_{H})/2}}{\left[\cos\left(\Omega_{C}\tau_{C}\right)-\frac{\alpha_{C}s\psi(0)-(\Gamma_{C}-\alpha_{C})/2}{\Omega_{C}}\sin\left(\Omega_{C}\tau_{C}\right)\right]\left[\cos\left(\Omega_{H}\tau_{H}\right)-\frac{\alpha_{H}(s-s\eta)\psi(\tau_{C}^{+})-(\Gamma_{H}-\alpha_{H})/2}{\Omega_{H}}\sin\left(\Omega_{H}\tau_{H}\right)\right]}\right\} ^{\frac{f_{n}}{2}}.
\end{equation}

\subsection{Self-consistency check of the joint generating function: fluctuation
theorem}

The fluctuation theorem for heat engines \citep{sinitsyn2011fluctuation}
provides a microscopic understanding of Carnot's theorem. For the
model presented here which never reaches thermal equilibrium with
the heat bath, the fluctuation theorem takes a different form. In
the following, we derive the fluctuation theorem for our finite-time
Brownian Carnot engine.

Suppose the engine operates between a hot bath at temperature $T_{H}$
and a cold bath at temperature $T_{C}$. Moreover, the the working
substance of the engine obeys the same Maxwell-Boltzmann distribution
at an effective temperature $\theta_{C}$ at both the beginning and
the end of the cycle (point A in Fig. 1(a)). The ratio of the probability
of a trajectory $X$ to that of its time reversal trajectory $\tilde{X}$
under the time reversed protocol is related to the entropy change
of the system and the heat functional of the forward trajectory \citep{seifert2012stochastic},
\begin{equation}
\frac{p[X]}{\tilde{p}[\tilde{X}]}=\exp(\Delta s^{bath}+\Delta s^{sys})=\exp\left(-\frac{Q_{H}}{T_{H}}-\frac{Q_{C}}{T_{C}}\right)\frac{\rho_{i}(x_{i})}{\rho_{f}(x_{f})}=\exp\left(-\frac{Q_{H}}{T_{H}}-\frac{Q_{C}}{T_{C}}+\frac{E_{f}-E_{i}}{\theta_{C}}\right),\label{eq:ratio_prob}
\end{equation}
where $Q_{H,C}$ denote the trajectory heat transferred from the bath
at temperature $T_{H,C}$ to the system, $\rho_{i,f}$ refer to the
initial and the final probability distribution of the system, and
$E_{i,f}$ denote the initial and the final stochastic energy of the
system in a cycle. According to the first law of thermodynamics along
a single trajectory, the stochastic energy change comprises the trajectory
work and heat, 
\begin{equation}
E_{f}-E_{i}=W+Q_{H}+Q_{C}.
\end{equation}
Accordingly, we can rewrite Eq. (\ref{eq:ratio_prob}) into 
\begin{equation}
p[X]\exp\left[Q_{H}\left(\frac{1}{T_{H}}-\frac{1}{\theta_{C}}\right)+Q_{C}\left(\frac{1}{T_{C}}-\frac{1}{\theta_{C}}\right)-\frac{W}{\theta_{C}}\right]=\tilde{p}[\tilde{X}].
\end{equation}
By taking ensemble average, we obtain the following fluctuation theorem
\begin{equation}
\left\langle \exp\left[Q_{H}\left(\frac{1}{T_{H}}-\frac{1}{\theta_{C}}\right)+Q_{C}\left(\frac{1}{T_{C}}-\frac{1}{\theta_{C}}\right)-\frac{W}{\theta_{C}}\right]\right\rangle =1.\label{eq:ft1}
\end{equation}
By choosing the beginning and the ending of a closed cycle at point
C {[}see Fig. 1(a){]}, we find another fluctuation theorem
\begin{equation}
\left\langle \exp\left[Q_{H}\left(\frac{1}{T_{H}}-\frac{1}{\theta_{H}}\right)+Q_{C}\left(\frac{1}{T_{C}}-\frac{1}{\theta_{H}}\right)-\frac{W}{\theta_{H}}\right]\right\rangle ^{\prime}=1,\label{eq:ft2}
\end{equation}
where the prime indicates the average is over a different ensemble
of trajectories from the former one. With Jensen's inequality $\exp\left\langle x\right\rangle \leq\left\langle \exp(x)\right\rangle $,
Carnot's theorem can be derived as a corollary of both fluctuation
theorems

\begin{equation}
\eta=-\frac{\left\langle W\right\rangle }{\left\langle Q_{H}\right\rangle }\leq1-\frac{T_{C}}{T_{H}}=\eta_{C},
\end{equation}
where we have used the conservation of the energy for a periodic steady
closed cycle $\left\langle W\right\rangle +\left\langle Q_{H}\right\rangle +\left\langle Q_{C}\right\rangle =\left\langle E_{f}\right\rangle -\left\langle E_{i}\right\rangle =0$.

Since the fluctuation theorem (\ref{eq:ft1}) can also be written
as $I_{\mathrm{cycle}}(1/T_{H}-1/\theta_{C},1/T_{C}-1/\theta_{C},-1/\theta_{C})=1$,
we can use it to check the validity of the analytical expression of
the joint generating function $I_{\mathrm{cycle}}(u_{H},u_{C},s)$.
In this case, the initial condition is $\tilde{\psi}_{0}=\theta_{C}$,
and a full cycle can be decomposed into four processes as in Fig.
1. The analytical expressions of the generating function associated
with these processes have been obtained previously. For a specific
set of parameters $u_{H}=1/T_{H}-1/\theta_{C},\:u_{C}=1/T_{C}-1/\theta_{C},\:s=-1/\theta_{C}$,
the auxiliary quantity $\psi(t)$ in the cycle can be obtained by
solving Eq. (\ref{eq:95-1})

\begin{equation}
\psi(t)=\begin{cases}
T_{C} & 0<t<\tau_{C}\\
T_{H} & \tau_{C}<t<\tau_{C}+\tau_{H},
\end{cases}
\end{equation}
and the shifted temperature-like variable is $\tilde{\psi}(t)\equiv\theta_{C}$
throughout the whole cycle. $\Omega_{C}$ and $\Omega_{H}$ are explicitly
$\Omega_{C}=i(\Gamma_{C}+\alpha_{C})/2$ and $\Omega_{H}=i(\Gamma_{H}+\alpha_{H})/2$.
Therefore, the joint generating function of each process becomes
\begin{align}
I_{\mathrm{I}} & =e^{-\frac{f_{n}}{2}\alpha_{C}\tau_{C}}\\
I_{\mathrm{II}} & =\left(1-\eta\right)^{\frac{f_{n}}{2}}\\
I_{\mathrm{III}} & =e^{-\frac{f_{n}}{2}\alpha_{H}\tau_{H}}\\
I_{\mathrm{IV}} & =\left(1-\eta\right)^{-\frac{f_{n}}{2}}.
\end{align}
Thus, we confirm the fluctuation theorem $I_{\mathrm{cycle}}(1/T_{H}-1/\theta_{C},1/T_{C}-1/\theta_{C},-1/\theta_{C})=1$
for heat engines. 

\subsection{Variances of the power and the efficiency }
\begin{widetext}
The derivatives of the generating function give the moments of work
distribution. The average work is 
\begin{equation}
\left\langle W\right\rangle =\frac{\partial}{\partial s}\left\langle \mathrm{e}^{sW}\right\rangle |_{s=0}=\frac{f_{n}}{2}(\theta_{H}-\theta_{C})\ln r,
\end{equation}
which can also be obtained from Eqs.(4,7) in the main text. The fluctuation
of the work output in a cycle can also be calculated from the generating
function
\begin{equation}
\left\langle \Delta W^{2}\right\rangle =\left\langle W^{2}\right\rangle -\left\langle W\right\rangle ^{2}=\left[\frac{\partial^{2}}{\partial s^{2}}\left\langle \mathrm{e}^{sW}\right\rangle -\left(\frac{\partial}{\partial s}\left\langle \mathrm{e}^{sW}\right\rangle \right)^{2}\right]|_{s=0}.
\end{equation}
The fluctuation of the power is defined as

\begin{equation}
\mathrm{Var}(P)\coloneqq\frac{\left\langle \Delta W^{2}\right\rangle }{(\tau_{C}+\tau_{H})^{2}},
\end{equation}
and the relative fluctuation of the power is given by
\begin{align}
\frac{\mathrm{Var}(P)}{\overline{P}^{2}}= & \frac{\left\langle \Delta W^{2}\right\rangle }{\left\langle W\right\rangle ^{2}}\nonumber \\
= & \frac{4}{f_{n}}\left[\frac{1}{\kappa_{C}\tau_{C}}(\frac{1-\eta}{\eta})^{2}+\frac{1}{\kappa_{H}\tau_{H}}\frac{1}{\eta^{2}}\right]+\frac{4}{f_{n}}\left[\frac{1-\mathrm{e}^{-\kappa_{H}\tau_{H}}}{\kappa_{H}^{2}\tau_{H}^{2}}\left(\frac{\kappa_{H}^{2}\tau_{H}^{2}}{(\ln r)^{2}}-\frac{1}{\eta^{2}}\right)-\frac{1-\mathrm{e}^{-\kappa_{C}\tau_{C}}}{\text{\ensuremath{\kappa}}_{C}^{2}\tau_{C}^{2}}(\frac{1-\eta}{\eta})^{2}\right]\nonumber \\
 & +\frac{4}{f_{n}}\frac{(1-\mathrm{e}^{-\kappa_{C}\tau_{C}})(1-\mathrm{e}^{-\kappa_{H}\tau_{H}})}{\kappa_{C}\tau_{C}\kappa_{H}\tau_{H}}\left(\frac{\kappa_{H}\tau_{H}}{\ln r}-\frac{1}{\eta}\right)\frac{1-\eta}{\eta},
\end{align}
where $\kappa_{C}=\Gamma_{C}-\ln r/\tau_{C}$ and $\kappa_{H}=\Gamma_{H}-\ln r/\tau_{H}$.
The fluctuation of efficiency is given by

\begin{align}
\mathrm{Var}(\text{\ensuremath{\zeta}}) & \text{\ensuremath{\coloneqq\frac{\left\langle (W+\eta Q_{H})^{2}\right\rangle }{\left\langle Q_{H}\right\rangle ^{2}}}}\nonumber \\
 & =\frac{4(1-\eta)^{2}}{f_{n}}\left[\frac{1}{\kappa_{C}\tau_{C}}+\frac{1}{\kappa_{H}\tau_{H}}\right]-\frac{4(1-\eta)^{2}}{f_{n}}\left[\frac{1-\mathrm{e^{-\kappa_{C}\tau_{C}}}}{\kappa_{C}^{2}\tau_{C}^{2}}+\frac{1-e^{-\kappa_{H}\tau_{H}}}{\kappa_{H}^{2}\tau_{H}^{2}}+\mathrm{-\frac{(1-e^{-\kappa_{C}\tau_{C}})(1-e^{-\kappa_{H}\tau_{H}})}{\kappa_{C}\tau_{C}\kappa_{H}\tau_{H}}}\right].
\end{align}
In order to achieve the maximum power, we should choose $\tau_{C}=\tau_{C}^{\mathrm{max}}$
and $\tau_{H}=\tau_{H}^{\mathrm{max}}$ according to Eq. (\ref{eq:max_time})
to realize the CA engine. As a result, the efficiency $\eta$ should
be replaced by $\eta_{CA}$.

Finally, we would like to analyze the scaling of the variance of the
power and the efficiency for the CA engine. When $\kappa_{C}\tau_{C}^{\mathrm{max}},\kappa_{H}\tau_{H}^{\mathrm{max}}\gg1$,
both variances $\mathrm{Var}(P)$ and $\mathrm{Var}(\zeta)$ decrease
inversely with the period of the cycle $\tau_{\mathrm{t}}\coloneqq\tau_{C}^{\mathrm{max}}+\tau_{H}^{\mathrm{max}}$
(note that for the CA engine, $\ln r$ is related to $\tau_{\mathrm{t}}$
by Eq. (\ref{eq:max_time})), that is
\begin{equation}
\mathrm{Var}(P)=\left\{ \frac{4}{f_{n}\eta_{CA}^{2}}\frac{\left[(1-\eta_{CA})^{2}+1/\delta\right]\left(1+\delta\right)}{\sqrt{\Gamma_{C}\Gamma_{H}}}\frac{1}{\tau_{\mathrm{t}}}+\mathcal{O}(\tau_{\mathrm{t}}^{-2})\right\} \overline{P}_{\mathrm{max}}^{2},
\end{equation}
and
\begin{equation}
\mathrm{Var}(\zeta)=\frac{4(1-\eta_{CA})^{2}}{f_{n}\sqrt{\Gamma_{C}\Gamma_{H}}}\frac{(1+\delta)^{2}}{\delta}\frac{1}{\tau_{\mathrm{t}}}+\mathcal{O}(\tau_{\mathrm{t}}^{-2}),
\end{equation}
where $\delta=\sqrt{\Gamma_{H}T_{H}/(\Gamma_{C}T_{C})}$.
\end{widetext}

\bibliographystyle{apsrev4-1}
\bibliography{ref,add_reference}